\documentclass[reprint,amsmath,amssymb,aps,prl]{revtex4-1}
\usepackage{graphicx}
\usepackage{amsmath,amssymb,braket,amsfonts}
\usepackage{dcolumn}
\usepackage{bm}

\begin{document}

\title{Randomness-induced quantum spin liquid behavior in the $s=$1/2 random $J_1$-$J_2$ Heisenberg antiferromagnet on the honeycomb lattice}
\author{Kazuki Uematsu} 
\author{Hikaru Kawamura}
  \email{kawamura@ess.sci.osaka-u.ac.jp}
\affiliation{
Department of Earth and Space Science, Graduate School of Science, Osaka University, Toyonaka, 560-0043, Japan}

\begin{abstract}
We investigate the ground-state and finite-temperature properties of the bond-random $s=1/2$ Heisenberg model on a honeycomb lattice with frustrated nearest- and next-nearest-neighbor antiferromagnetic interactions, $J_1$ and $J_2$, by the exact diagonalization and the Hams--de Raedt methods. The ground-state phase diagram of the model is constructed in the randomness versus the frustration ($J_2/J_1$) plane, with the aim of clarifying the effects of randomness and frustration in stabilizing a variety of distinct phases. We find that the randomness induces the gapless quantum spin liquid (QSL)-like state, the random-singlet state, in a wide range of parameter space. The observed robustness of the random-singlet state suggests that the gapless QSL-like behaviors might be realized in a wide class of frustrated quantum magnets possessing a certain amount of randomness or inhomogeneity, without fine-tuning the interaction parameters. Possible implications to recent experiments on the honeycomb-lattice magnets Ba$_3$CuSb$_2$O$_9$ and 6HB-Ba$_3$NiSb$_2$O$_9$ exhibiting the gapless QSL-like behaviors are discussed.
\end{abstract}

\maketitle

\section{Introduction}
\label{sec:intro}

The quantum spin liquid (QSL) state without any spontaneously broken Hamiltonian symmetry, which accompanies no magnetic long-range order (LRO) down to low temperatures, has long received much attention.\cite{QSL} For the realization of such QSL state, geometrical frustration is considered to be essential, and frustrated magnets have been the main target of the quest for QSL materials. In particular, the $s=1/2$ organic triangular-lattice salts $\kappa$-(ET)$_2$Cu$_2$(CN)$_3$,\cite{ETsalt, ETsaltCv,ET-Matsuda,ET-ChargeGlass,ET-Sasaki} EtMe$_3$Sb[Pd(dmit)$_2$]$_2$, \cite{dmitsalt,dmitsalt-Matsuda,dmitsaltCv,dmit-ChargeGlass} and more recently $\kappa$-H$_3$(Cat-EDT-TTF)$_2$ \cite{Isono1,Isono2,Ueda} were reported to exhibit the QSL-like behaviors down to very low temperatures. The QSL states of these organic salts commonly exhibit gapless (or nearly gapless) behaviors characterized by, {\it e.g.\/}, the low-temperature specific heat linear (or almost linear) in the absolute temperature $T$. \cite{ETsaltCv,ET-Matsuda,dmitsalt-Matsuda,dmitsaltCv} Another well-studied candidate of the QSL might be the $s=1/2$ kagome-lattice inorganic compound herbertsmithite ZnCu$_3$(OH)$_6$Cl$_2$. \cite{Shores,Helton,Olariu,Freedman,Han,Imai} This kagome material was also reported to exhibit gapless QSL-like behaviors, \cite{Helton,Olariu,Han} while a recent NMR study showed a nonzero spin gap. \cite{Imai}

 Despite such recent experimental progress, the true origin of the experimentally observed QSL-like behaviors still remains not fully understood and is under hot debate. In many theoretical studies, it has been assumed that the system is sufficiently clean so that the possible effect of randomness or inhomogeneity is negligible and unimportant. Meanwhile, one of the present authors (H.K.) and collaborators have claimed that the QSL-like behaviors recently observed in triangular-lattice organic salts and kagome-lattice herbertsmithite might be the randomness-induced one, the random-singlet state. \cite{Watanabe,Kawamura,Shimokawa} The advocated random-singlet state is a gapless QSL-like state where spin singlets of varying strengths are formed in a hierarchical manner on the background of randomly distributed exchange interactions $J_{ij}$. The state may also be regarded as a sort of ``Anderson-localized resonating valence bond (RVB) state''. Indeed, it was demonstrated that the random-singlet state exhibited the $T$-linear specific heat and the gapless susceptibility with an intrinsic Curie tail, accompanied by the gapless and broad features of the dynamical spin structure factor. \cite{Watanabe,Kawamura,Shimokawa}

 The source of the randomness or inhomogeneity in real materials could be various. For triangular organic salts, in Ref. \cite{Watanabe}, it was argued that the {\it effective\/} randomness or inhomogeneity might be self-generated  for the spin degrees of freedom via the coupling between the spin and the charge, owing to the charge order inherent to these compounds and the associated slowing down of the electric-polarization degrees of freedom at each dimer molecule. Indeed, there reported some experimental evidence of such spatial inhomogeneity in the charge and spin distributions in triangular organic salts. \cite{ET-Sasaki,Kawamoto,Shimizu,Nakajima} For the kagome herbertsmithite, in Ref. \cite{Kawamura}, it was suggested that the random Jahn--Teller (JT) distortion of the [Cu(OH)$_6$]$^{4-}$ octahedra driven by the random substitution of magnetic Cu$^{2+}$ for nonmagnetic ${\rm Zn^{2+}}$ on the adjacent triangular layer \cite{Freedman} might give rise to the random modification of the exchange paths between various neighboring magnetic Cu$^{2+}$ pairs on the kagome layer, leading to the random modulation of the exchange couplings between various neighboring Cu$^{2+}$ pairs on the kagome layer.

The possible important role played by the randomness or inhomogeneity has also been reported in other triangular and kagome magnets as well. One example might be an inorganic triangular antiferromagnet Cs$_2$Cu(Br$_{1-x}$Cl$_x$)$_4$, a mixed crystal of  Cs$_2$CuBr$_4$ and Cs$_2$CuCl$_4$. This random magnet was observed not to exhibit the magnetic LRO nor the spin-glass freezing down to very low temperatures in a certain range of $x$. \cite{Ono} Another intriguing example might be the triangular-lattice organic salt $\kappa$-(ET)$_2$Cu[N(CN)$_2$]Cl, which exhibits the standard AF LRO, but exhibits the QSL-like behaviors when the randomness is artificially introduced by X-ray irradiation. \cite{Furukawa} Another example might be the $s=1/2$ kagome-like Heisenberg AF, ZnCu$_3$(OH)$_6$SO$_4$, where the spin-1/2 Cu$^{2+}$ is located on the corrugated kagome plane. This kagome material contains a significant amount of randomness and was found to exhibit gapless QSL-like behaviors characterized by the $T$-linear specific heat. \cite{Li,Zorko,Zorko2} Thus, there now emerge growing experimental lines of evidence that the randomness-induced QSL-like state exists in nature.

\begin{figure}
  \begin{center}
    \includegraphics[clip,width=0.5\hsize]{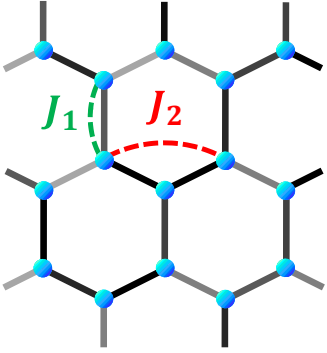}\par
    \caption{Illustration of the honeycomb lattice, and the nearest- and next-nearest-neighbor interactions $J_1$ and $J_2$.}
    \label{fig:honeycomb}
  \end{center}
\end{figure}

 From a theoretical viewpoint, one might naturally ask whether the combined effect of randomness and quantum fluctuations is sufficient to cause the QSL-like behaviors. A counter-example of this was reported for the case of the random-bond $s=1/2$ AF Heisenberg model on the square lattice. In contrast to the frustrated triangular- or kagome-lattice counterparts obeying the same form of randomness distribution, this unfrustrated model persists to exhibit the AF LRO up to the maximal randomness. \cite{RandomSquare1,Watanabe} This observation certainly suggests that the frustration also plays an important role in realizing the QSL-like behaviors. 

 Under such circumstances, it is important to clarify the role of frustration along with that of randomness in realizing the random-singlet state in quantum magnets. In this paper, we wish to undertake such a study by investigating the properties of a model where the strengths of both randomness and frustration can independently be tuned. For this purpose, we choose the $s=1/2$ honeycomb-lattice Heisenberg model with the competing AF nearest- and next-nearest-neighbor interactions $J_1$ and $J_2$, as shown in Fig. \ref{fig:honeycomb}. The honeycomb lattice is bipartite so that the $J_1$-only model is unfrustrated. Frustration is introduced via the competition between $J_1$ and $J_2$, the ratio $J_2/J_1$ representing the extent of frustration.  The choice of the honeycomb lattice is motivated by the fact that the honeycomb lattice has only three nearest neighbors, a minimal number among various 2D lattices, and is subject to the enhanced effects of fluctuations possibly destroying the magnetic LRO.
\begin{figure}
  \begin{center}
    \includegraphics[clip,width=0.99\hsize]{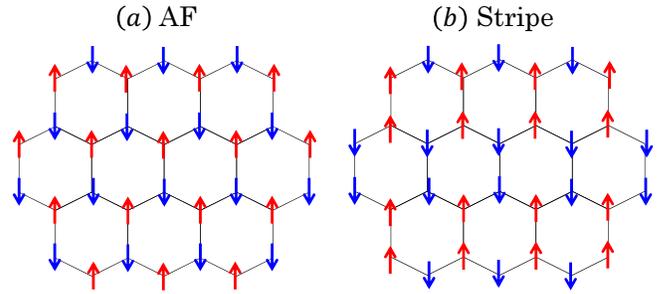}\par
    \caption{Candidates of the magnetically ordered states of the $J_1$-$J_2$  Heisenberg antiferromagnet on the honeycomb lattice; (a) two-sublattice AF state, and (b) stripe-ordered state.}
    \label{fig:magneticGSsketch}
  \end{center}
\end{figure}

\begin{figure}
  \begin{center}
    \includegraphics[clip,width=0.99\hsize]{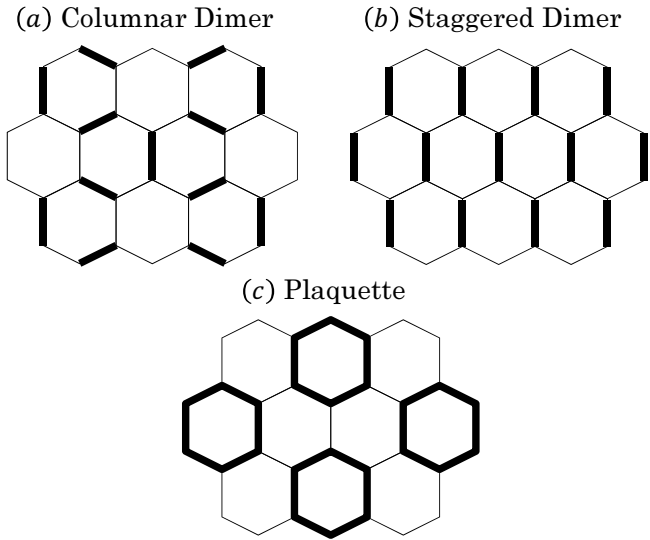}
    \caption{Candidates of the nonmagnetic states of the $J_1$-$J_2$ Heisenberg antiferromagnet on the honeycomb lattice; (a) columnar-dimer state, (b) staggered-dimer state, and (c) plaquette state.}
    \label{fig:GSsketch}
  \end{center}
\end{figure}

 In fact, the low-temperature properties of the regular $s=1/2$ $J_1$-$J_2$ Heisenberg model on the honeycomb lattice have attracted much interest. \cite{Fouet,Mulder,Mosadeq,Lauchli,White,Ganesh,Gong} In particular, the ground-state phase diagram of the model was extensively studied as a function of $J_2$ by various methods, including the exact diagonalization (ED) \cite{Fouet,Mosadeq,Lauchli} and the density matrix renormalization group (DMRG) \cite{White,Ganesh,Gong} methods. When $J_2$ is small, the ground state is the standard two-sublattice AF state as illustrated in Fig. \ref{fig:magneticGSsketch}(a). When the strength of frustration exceeds the critical value $J_{2c}$, the AF state is destabilized and QSL-like states emerge. \cite{Fouet,Mosadeq, Lauchli,White, Ganesh,Gong} In fact, the critical $J_2$ value of the AF instability is around $0.2-0.25$, which is greater than the classical counterpart $J_{2c}=1/6$. Beyond $J_{2c}$, several types of nonmagnetic states have been discussed in the literature, some of which are illustrated in Fig. \ref{fig:GSsketch}. They include the columnar-dimer state [Fig. \ref{fig:GSsketch}(a)], which breaks both the lattice translational and the lattice rotational symmetries, the staggered-dimer state [Fig. \ref{fig:GSsketch}(b)], which keeps the lattice translational symmetry but breaks the lattice rotational symmetry,\cite{Fouet, Mosadeq, White, Ganesh} and the plaquette state [Fig. \ref{fig:GSsketch}(c)], which keeps the lattice rotational symmetry but breaks the lattice translational symmetry. \cite{Fouet,Mosadeq,Lauchli,White,Ganesh,Gong} These states are all variants of the valence-bond crystal (VBC) state with a nonzero spin gap. In addition, the possible occurrence of the $Z_2$ spin-liquid state was also discussed (see, {\it e.g.}, Ref. \cite{Gong} and references cited therein). The magnetically ordered state such as the stripe-ordered state illustrated in Fig. \ref{fig:magneticGSsketch}(b) was also invoked as a ground state of the larger $J_2$ region. {\cite{Lauchli}} Our present study addresses the issue of what happens to these magnetic and nonmagnetic states of the regular model if one introduces the randomness.

 Experimentally, the magnetic ordering of honeycomb-lattice AFs has attracted much recent interest. Owing to the bipartite nature of the lattice, some of them exhibit the standard AF order,\cite{Cu5SbO6} whereas some are reported not to exhibit the magnetic order.\cite{Azuma,Matsuda,Zhou,Nakatsuji,Mendels-Cu,Ishiguro,Do,Hagiwara,Katayama,Cheng,Darie,Mendels-Ni} Even if one sets aside the dimerized nonmagnetic state with a finite spin gap, occasionally induced by the uniform JT structural distortion of the lattice, interesting spin-liquid-like behaviors have been reported in several honeycomb magnets.

 One example might be Bi$_3$Mn$_4$O$_{12}$(NO$_3$) (BMNO), an insulating $s=3/2$ system. This material remains paramagnetic down to low temperatures in zero field with a small spin correlation length, but exhibits the AF order upon applying even weak fields. \cite{Azuma,Matsuda} Since the Mn$^{4+}$ spin is 3/2 here, the spin-liquid-like behaviors of BMNO might be expected to be essentially of classical origin. Indeed, the spin-liquid-like behavior and the field-induced AF of this material have successfully been explained within the classical (or semi-classical) picture as the ``ring liquid'' and the ``pancake liquid'' arising from the characteristic ring-like degeneracy in the wavevector space. \cite{Okumura} The manner by which this degeneracy is lifted by fluctuations via the `order-from-disorder' mechanism leading to the magnetic ordering was also studied. \cite{Okumura,Mulder} In particular, in Ref. \cite{Okumura}, it was argued that, close to the AF instability, which was equal to $J_{2c}=1/6$ in the classical limit, the energy scale of this order from disorder could be very small such that the classical spin-liquid state (the ring liquid or the pancake liquid) might be stabilized down to very low temperatures.

\begin{figure}
  \begin{center}
    \includegraphics[clip,width=0.99\hsize]{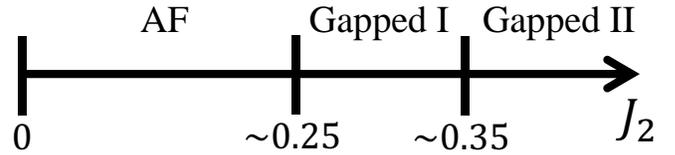}
    \caption{Ground-state phase diagram of the $s=1/2$ regular ($\Delta=0$) $J_1$-$J_2$ Heisenberg model on the honeycomb lattice. `AF' represents the standard two-sublattice antiferromagnetic state, while `Gapped I' and `Gapped II' represent the nonmagnetic states with a finite spin gap. The gapped I and II states are likely to be the plaquette state and the staggered-dimer state, respectively.}
    \label{fig:del00phase}
  \end{center}
\end{figure}

 A different type of spin-liquid-like behavior, likely to be of quantum origin, has recently been reported in the honeycomb-lattice-based compounds Ba$_3$CuSb$_2$O$_9$, \cite{Zhou,Nakatsuji,Mendels-Cu,Ishiguro,Do,Hagiwara,Katayama} and 6HB-Ba$_3$NiSb$_2$O$_9$. \cite{Cheng,Darie,Mendels-Ni} The former Cu$^{2+}$ compound Ba$_3$CuSb$_2$O$_9$  has $s=1/2$ forming a decorated honeycomb lattice, with a significant amount of randomness associated with the Cu-Sb `dumbbell' orientation. \cite{Nakatsuji} Stoichiometric samples tend to be hexagonal, whereas off-stoichiometric samples tend to be orthorhombic with a static JT transition occurring at a higher temperature $T\simeq 200$K. \cite{Nakatsuji,Hagiwara} Although the orthorhombic sample was reported to exhibit the spin freezing at a low temperature $T_g\simeq 110$ mK, the hexagonal sample does not order magnetically down to very low temperatures. \cite{Nakatsuji} For such hexagonal samples, in Ref. \cite{Nakatsuji}, two different scenarios were suggested. In one, the spin and orbital degrees of freedom are entangled and fluctuate together down to low temperatures forming a sort of spin-orbital liquid (the dynamical JT effect), \cite{Nakatsuji,Ishiguro,Do,Hagiwara,Katayama} while, in the other, the static random JT distortion might lead to the `random-singlet state'. \cite{Mendels-Cu} Many of the recent studies seem to point to the former scenario, while the NMR study of Ref. \cite{Mendels-Cu} favored the second scenario, {\it i.e.\/}, the `random-singlet state' characterized by the two kinds of gaps, $\Delta_1$ and $\Delta_2$, $\Delta_1$ ($>\Delta_2$) being in the order of the main exchange interaction and $\Delta_2$ induced by applied fields. For the orthorhombic samples, the lattice symmetry is lowered from hexagonal to orthorhombic even macroscopically at temperatures below the static hexagonal-to-orthorhombic JT transition at 200 K, which might affect the underlying low-temperature physics. Indeed, the spin-glass freezing was reported at a low temperature of 110 mK in the orthorhombic sample. \cite{Nakatsuji}

 Hence, the situation for Ba$_3$CuSb$_2$O$_9$ still remains not totally clear, including the question on how the random-singlet state identified for the bond-random triangular and kagome models, \cite{Watanabe,Kawamura,Shimokawa} {\it i.e.\/}, the randomness-induced gapless QSL-like state, is related or unrelated to the `random-singlet state' discussed in the literature for the honeycomb-lattice-based AFs Ba$_3$CuSb$_2$O$_9$ both in the hexagonal and orthorhombic samples.

 By contrast, the latter Ni$^{2+}$ compound 6HB-Ba$_3$NiSb$_2$O$_9$ synthesized in high fields and at high temperatures has $s=1$. \cite{Cheng,Darie,Mendels-Ni} Recent analysis has revealed that the structure of this compound is likely to be trigonal rather than hexagonal, forming a honeycomb lattice made up of triangular bilayers with a significant amount of randomness contained, where the nearest- and the next-nearest-neighbor interactions $J_1$ and $J_2$ arise from the alternating arrangement of the Ni-Sb dumbbell on adjacent triangular layers. \cite{Darie,Mendels-Ni} In this Ni$^{2+}$ compound, the orbital degrees of freedom are absent unlike the Cu$^{2+}$ counterpart, yet the system exhibits the QSL-like behavior characterized by the $T$-linear specific heat. One might wonder if the QSL-like behavior of this material might have any connection with the randomness-induced gapless QSL-like states of triangular and kagome AFs. \cite{Watanabe,Kawamura,Shimokawa} This possible connection provides another motivation of our present study.

 The structure of this article is as follows. In sect. \ref{sec:modelmethod}, we introduce our model on the honeycomb lattice, the random-bond $s=1/2$ $J_1$-$J_2$ Heisenberg model, and explain the computational method employed. In sect. \ref{sec:del00}, we summarize the ground-state properties of the corresponding {\it regular\/} model on the basis of the previous works and some new data of our own. This sect. serves the basis of our study on the random model in the following sect. Sections \ref{sec:del--} and \ref{sec:finitemp} are the main parts of the present paper. The ground-state properties of the {\it random\/} model are studied in sect. \ref{sec:del--}. The ground-state phase diagram is constructed in the frustration ($J_2/J_1$) versus the randomness (the parameter $\Delta$ is defined below) plane, and the properties of each phase are clarified. The finite-temperature properties of the model are studied in sect. \ref{sec:finitemp}. Section \ref{sec:summary} is devoted to summary and discussion.

\begin{figure*}[t]
  \begin{center}
    \begin{tabular}{c}
      \begin{minipage}{0.3\hsize}
        \begin{center}
          \includegraphics[clip,width=0.99\hsize]{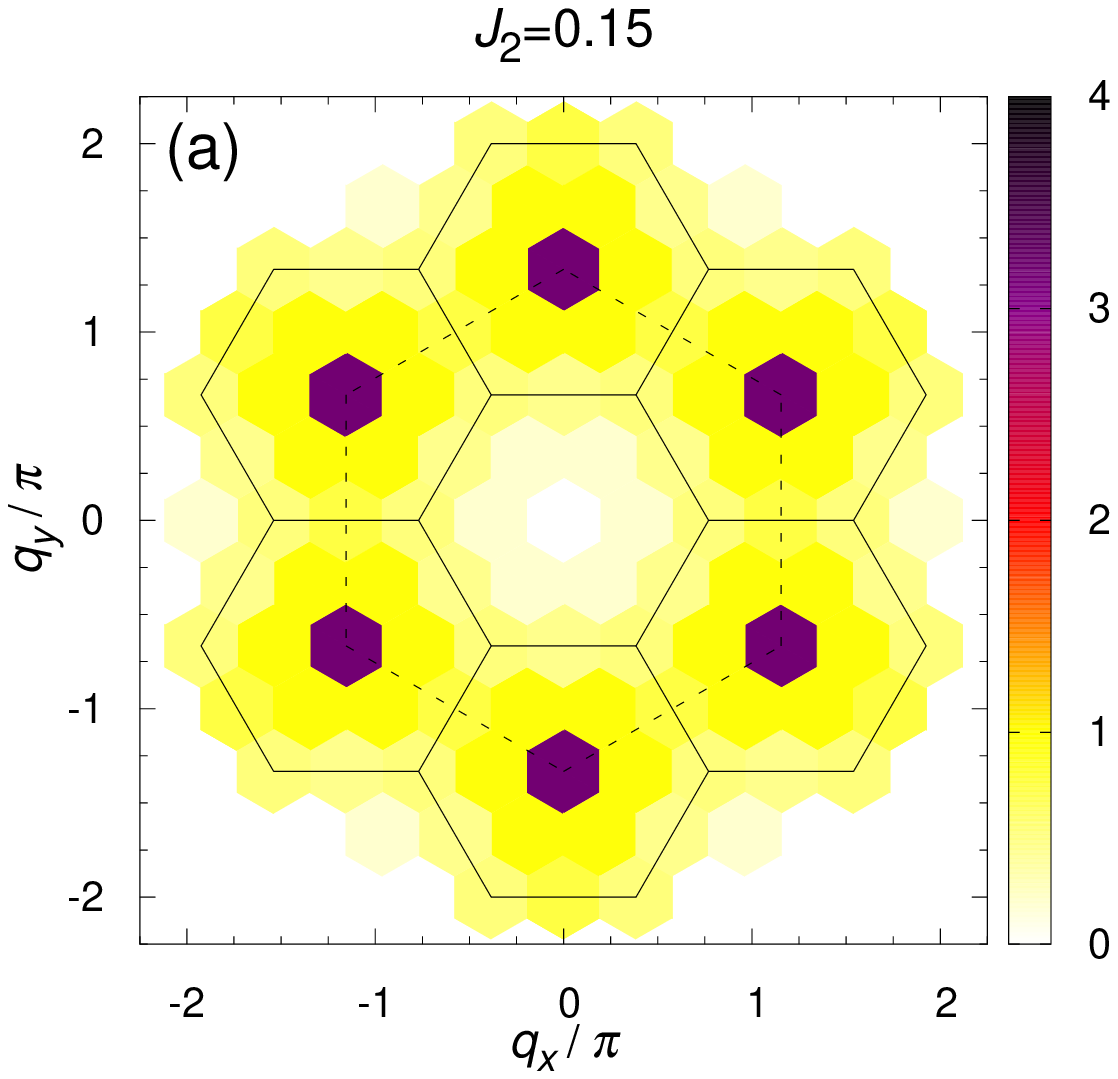}
       \end{center}
      \end{minipage}
      \begin{minipage}{0.3\hsize}
        \begin{center}
          \includegraphics[clip,width=0.99\hsize]{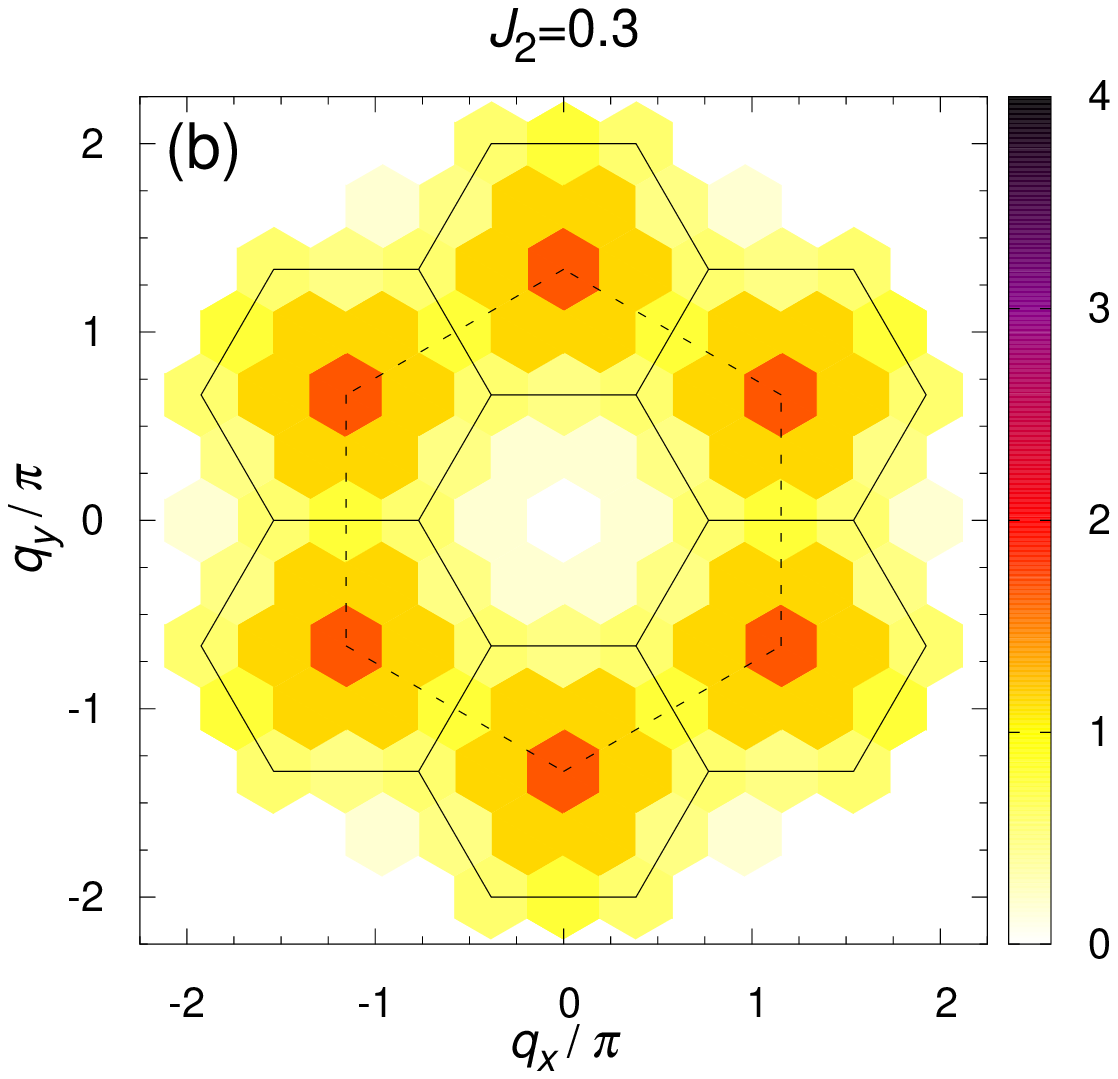}
        \end{center}
      \end{minipage}
      \begin{minipage}{0.3\hsize}
        \begin{center}
          \includegraphics[clip,width=0.99\hsize]{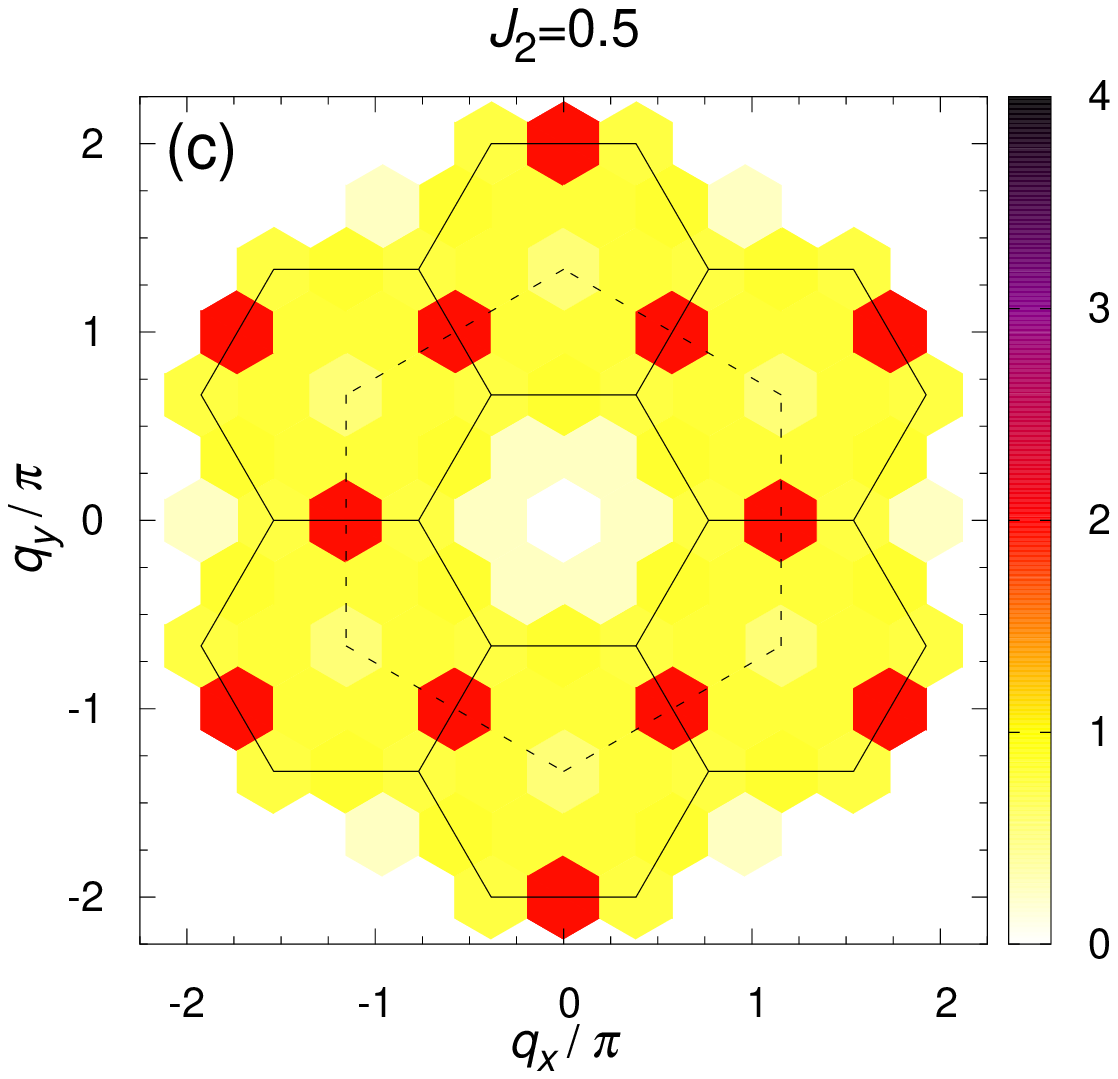}
        \end{center}
      \end{minipage}
    \end{tabular}
    \caption{Intensity plots of the static spin structure factor $S_{\bm q}$ of the regular model at (a) $J_2=0.15$, (b) $J_2=0.3$, and (c) $J_2=0.5$. The lattice size is  $N=24$. The solid line shows the boundary of the first Brillouin zone of the triangular lattice composing a sublattice of the original honeycomb lattice. The dashed line shows the boundary of the extended Brillouin zone.}
    \label{fig:SSF-regular}
  \end{center}
\end{figure*}

\section{Model and method}
\label{sec:modelmethod}

We consider the bond-random $s=1/2$ isotropic Heisenberg model on the honeycomb lattice with the AF nearest-neighbor and next-nearest-neighbor interactions $J_1$ and $J_2$. The Hamiltonian is given by
\begin{align}
\mathcal{H}=J_1\sum_{\Braket{i,j}}j_{ij}{\bm S}_i\cdot{\bm S}_j +
J_2\sum_{\Braket{\Braket{i,j}}} j_{ij}{\bm S}_i\cdot{\bm S}_j,
\label{eq:hamiltonian}
\end{align}
where ${\bm S}_i=(S_i^x, S_i^y, S_i^z)$ is an $s=1/2$ spin operator at the $i$-th site on the honeycomb lattice, the sums $\Braket{i,j}$ and $\Braket{\Braket{i,j}}$ are taken over all nearest-neighbor and next-nearest-neighbor pairs on the lattice where periodic boundary conditions are assumed, while $j_{ij}\ge 0$ is the random variable obeying the bond-independent uniform distribution between $[1-\Delta, 1+\Delta]$ with $0\leq \Delta\leq 1$. Hereafter, we put $J_1=1$ and $J_2/J_1=J_2>0$. Then, the parameter $J_2$ represents the degree of frustration, because the frustration in the present model is exclusively borne by the competition between the nearest- and next-nearest-neighbor interactions. Our present choice of the bond-independent uniform distribution for $j_{ij}$ is just for simplicity, whereas, in real materials, the distribution could be more complex and correlated. The parameter $\Delta$ represents the extent of the randomness: $\Delta=0$ corresponds to the regular case and $\Delta=1$ to the maximally random case. Again, the extent of the randmness $\Delta$ is taken to be common between $J_1$ and $J_2$ just for simplicity. By tuning the parameters $\Delta$ and $J_2$, we can control the degrees of both the randomness and the frustration independently.

 The ground-state properties of the model are computed by the ED Lanczos method. We treat finite-size clusters with the total number of spins $N$ up to $N\leq32$ (all even-$N$ samples with $8\leq N\leq 32$), with periodic boundary conditions being applied. All clusters studied are commensurate with the two-sublattice AF order illustrated in Fig. \ref{fig:magneticGSsketch}(a), and with the staggered-dimer order illustrated in Fig. \ref{fig:GSsketch}(b). The clusters of $N=8$, 12, 16, 20, 24, 28, and 32 are commensurate with the stripe order of Fig. \ref{fig:magneticGSsketch}(b), among which $N=8$, 24, and 32 possess the threefold rotational symmetry of the bulk honeycomb lattice. (More generally, the clusters of $N=8$, 14, 18, 24, 26, and 32 possess the threefold rotational symmetry of the bulk honeycomb lattice.) The clusters of $N=12$, 18, 24, and 30 are commensurate with the columnar-dimer and the plaquette orders shown respectively in [Figs. \ref{fig:GSsketch}(a) and \ref{fig:GSsketch}(c)].  

 The numbers of independent bond realizations, $N_s$, used in the configurational or sample average are $N_s=100$, 50, 25, 16, and 10 for $N=8-24$, 26, 28, 30, and 32 for the order parameter, the spin gap, and the static spin structure factor,  whereas $N_s=100$, 100, and 25 for $N=18,24$, and 32 for the dynamical spin structure factor, respectively. Error bars are estimated from sample-to-sample fluctuations.

\begin{figure}
  \begin{center}
    \includegraphics[width=7.5cm]{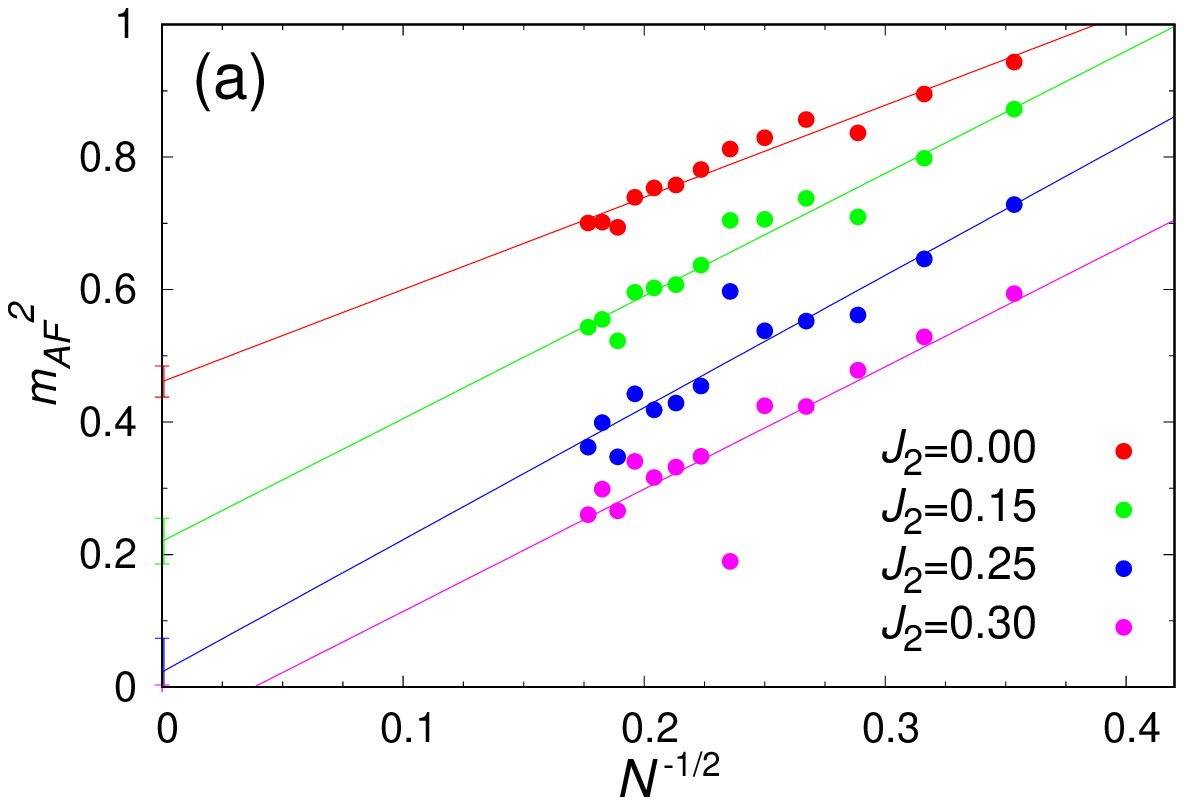}
    \includegraphics[width=7.5cm]{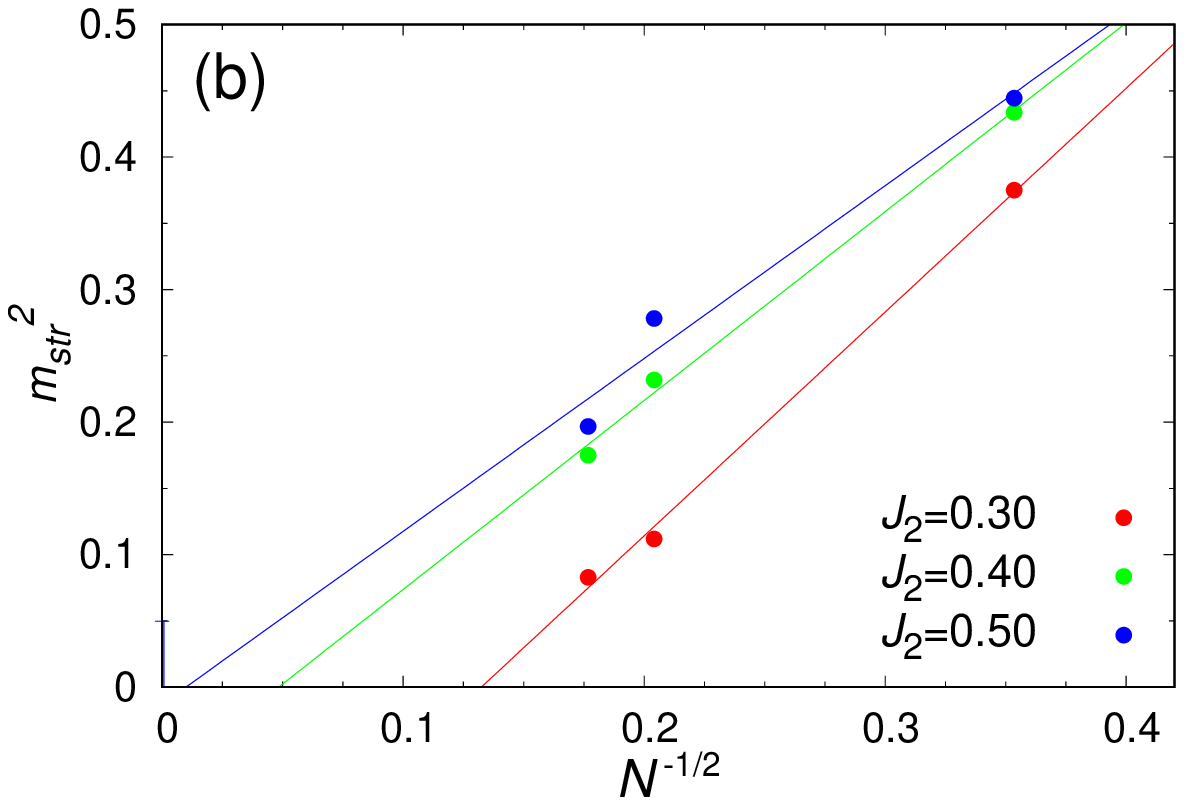}
    \caption{Size dependence of the magnetic order parameters of the regular model; (a) squared sublattice magnetization of the standard two-sublattice AF order $m_{AF}^2$, and (b) squared sublattice magnetization of the stripe order $m_{str}^2$, plotted versus $1/\sqrt{N}$ for various values of $J_2$.}
    \label{fig:del00magnet-order}
  \end{center}
\end{figure}

 The finite-temperature properties are computed by the Hams--de Raedt method, \cite{HamsRaedt} where the thermal average is replaced by the average over a few pure states produced via the imaginary time-evolution of initial vectors. The method enables us to calculate various finite-temperature properties at nearly the same computational cost as the Lanczos method. Our finite-temperature computation is performed for the size $N=24$, where the averaging is made over 30 initial vectors and 50 independent bond realizations. Error bars of physical quantities are estimated from the scattering over both samples and initial states.

\section{Ground-state properties of regular model}
\label{sec:del00}

 In this section, we present the ground-state properties of the regular $(\Delta=0)$ $J_1$-$J_2$ model on the honeycomb lattice. There already exist some ED works on this model, \cite{Fouet, Mosadeq, Lauchli} even to sizes larger than our largest one $N=32$. Even so, we feel that presenting some of our data in the form appropriate for later comparison with the random model might be useful.

\subsection{Possible phases and phase diagram}

 First, we summarize in Fig. \ref{fig:del00phase} the ground-state phase diagram of the regular $J_1-J_2$ honeycomb model in the region of $0\leq J_2\leq 0.5$, obtained on the basis of previous calculations and our present ED calculations. Three distinct phases appear, {\it i.e.\/}, the two-sublattice AF phase, and the nonmagnetic gapped I and II phases. For the smaller $J_2$ region of $J_2\lesssim 0.25$, the standard two-sublattice AF state is stabilized owing to the bipartite nature of the honeycomb lattice. When $J_2$ is increased beyond a critical value of $J_{2c}\simeq 0.25$, the nonmagnetic gapped state, the gapped I state, appears. The state is magnetically disordered with a finite spin gap. As a candidate of this gapped I phase, many previous studies have pointed to the plaquette phase, although no complete consensus has emerged. \cite{Fouet, Mosadeq,Lauchli} The plaquette state maintains the threefold lattice rotational symmetry but breaks the translational symmetry as can be seen from Fig. \ref{fig:GSsketch}(c). Another candidate of the gapped I phase might be the $Z_2$ spin liquid phase. \cite{Gong} Most studies reported the critical $J_2$ value of around $0.2-0.25$. \cite{Mosadeq, Lauchli,White, Ganesh} When $J_2$ is further increased beyond $J_2=0.35$-0.4, there occurs a transition from the gapped I phase to another gapped phase (gapped II phase), where the state is still magnetically disordered with a finite spin gap. As a candidate of this gapped II phase, many previous studies have pointed to the staggered-dimer state, \cite{Fouet,Mosadeq,White,Ganesh} also called the lattice nematic state.\cite{Mulder}  The staggered-dimer state maintains the translational symmetry, but breaks the threefold rotational symmetry as can be seen from Fig. \ref{fig:GSsketch}(b). Another candidate of the gapped II phase might be the magnetic stripe phase shown in Fig. \ref{fig:magneticGSsketch}(b). \cite{Lauchli}
\begin{figure*}[t]
  \begin{center}
    \includegraphics[clip,width=0.99\hsize]{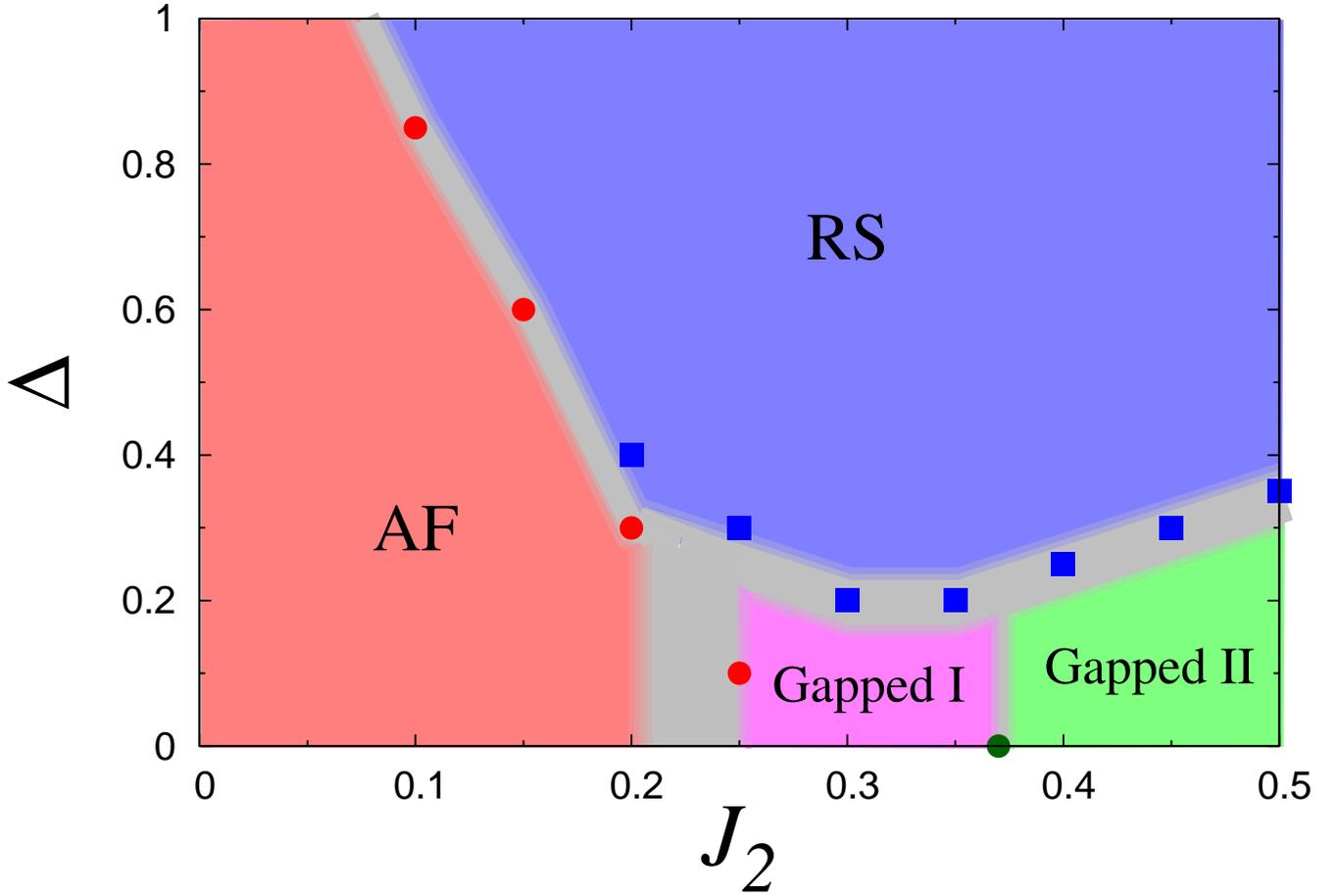}
    \caption{Ground-state phase diagram of the $s=1/2$ random ($\Delta\neq0$) $J_1$-$J_2$ Heisenberg model on the honeycomb lattice in the frustration ($J_2$) versus the randomness ($\Delta$) plane. `AF' and 'RS' represent the antiferromagnetic state and the random-singlet state, while `Gapped I' and `Gapped II' represent the nonmagnetic states with a finite spin gap, each of which is likely to be the plaquette, and the staggered-dimer states, respectively. The red points denote the transition points estimated from the AF order parameter, while the blue and green points denote those estimated from the spin gap and the static spin structure factor.} 
    \label{fig:phasediagram-random}
  \end{center}
\end{figure*}

 We perform some ED calculations to obtain additional information on the ground-state properties of the regular model, and obtain the results basically supporting the phase diagram of Fig. \ref{fig:del00phase}. In Fig. \ref{fig:SSF-regular}, we show the computed ground-state spin structure factor $S_{\bm q}$ defined by
\begin{align}
S_{\bm q}&=\frac{1}{N}\left[ \Braket{ |{\bm S}_{\bm q}|^2}\right]_J \nonumber \\
&=\frac{1}{N} \left[ \sum_{i,j} \Braket{{\bm S}_i\cdot{\bm S}_j}
  \cos{\left({\bm q}\cdot\left({\bm r}_i-{\bm r}_j\right)\right)}
  \right]_J,
\label{eq:SSF}
\end{align}
where ${\bm S}_{\bm q}=\sum_j {\bm S}_j e^{i{\bm q}\cdot{\bm r}_j}$ is the Fourier transform of the spin operator, ${\bm r}_j$ is the position vector at the site $j$, ${\bm q}$ is the wavevector, while $\langle \cdots \rangle$ and $[\cdots]_J$ represent the ground-state expectation value (or the thermal average at finite temperatures) and the configurational average over $J_{ij}$ realizations. The $J_2$ values are (a) $J_2=0.15$, (b) 0.3, and (c) 0.5.

 For the $J_2$-values smaller than $J_{2c}\simeq 0.25$, a rather sharp peak corresponding to the standard two-sublattice AF order appears  at the $\Gamma$ point located at ${\bm q}=\pi(\frac{2}{\sqrt{3}}, \frac{4}{3})$, where the length unit is taken here to be the nearest-neighbor distance of the honeycomb lattice. An example for the case of $J_2=0.15$ is shown in  Fig. \ref{fig:SSF-regular}(a). As $J_2$ is increased to $J_2=0.3$, the AF peak at the $\Gamma$ point becomess broadened, but without any other peaks appearing. The $J_2=0.3$ point lies in the gapped I phase in the phase diagram of Fig. \ref{fig:del00phase} so that the broad peak observed here is likely to be associated with the magnetic short-range order (SRO), which will be confirmed by our data of the AF order parameter shown below. Interestingly, at a still larger value of $J_2=0.5$ shown in Fig. \ref{fig:SSF-regular}(c) lying in the gapped II phase in the phase diagram of Fig. \ref{fig:del00phase}, the peak remains broad, but its position moves from the $\Gamma$ point to the M points located at ${\bm q}=\pi(\frac{2}{\sqrt{3}}, 0)$, $\pi (\frac{1}{\sqrt{3}},3)$, and $\pi (\frac{1}{\sqrt{3}}, -3)$. From the lattice symmetry, there exist three equivalent but independent $M$ points, which, in terms of the ordering pattern, corresponds to the three distinct types of the stripe order \cite{Lauchli} illustrated in Fig. \ref{fig:magneticGSsketch}(b). The observed broadness of the M-point peaks indicates that the stripe order remains to be a SRO here, not a true LRO, which will be confirmed also by our data of the stripe order parameter shown in Fig. \ref{fig:del00magnet-order}(b).

\begin{figure}
  \begin{center}
    \includegraphics[width=7.5cm]{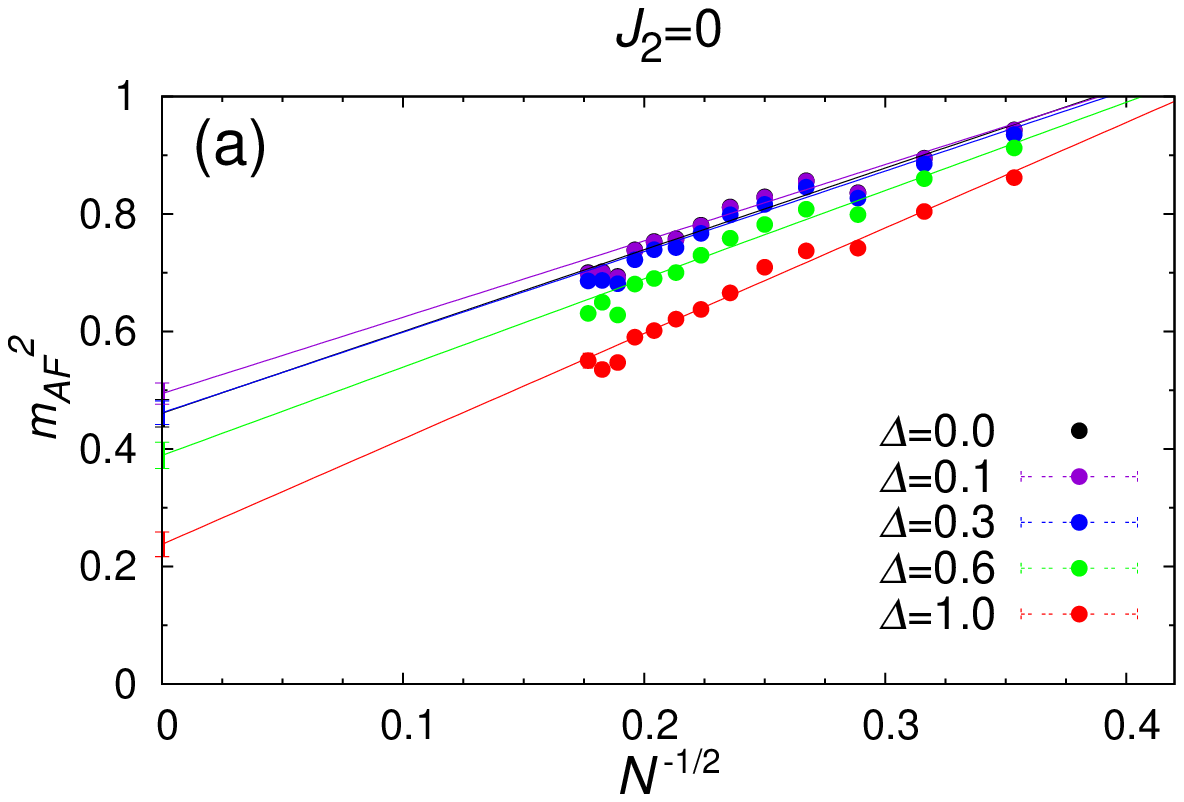}\par
    \includegraphics[width=7.5cm]{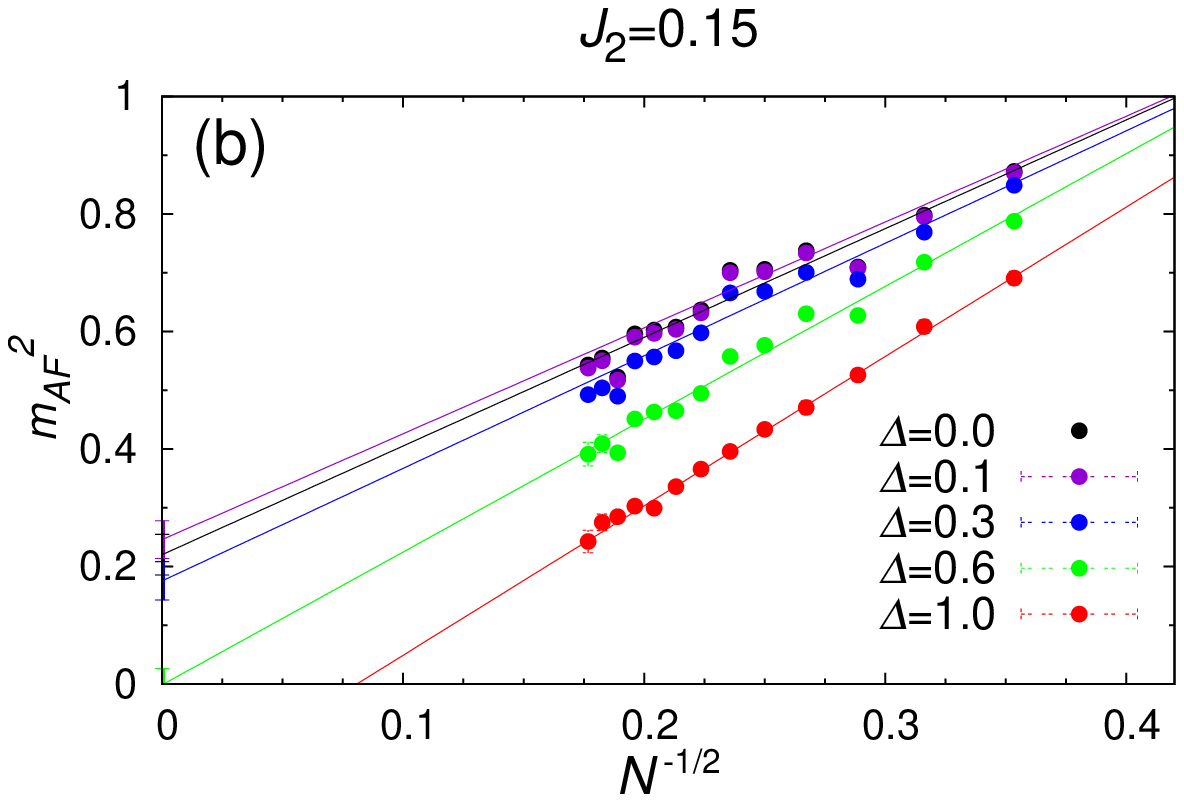}\par
    \caption{Squared sublattice magnetization of the AF order $m_{AF}^2$ of the random model plotted versus $1/\sqrt{N}$ for various values of $\Delta$, at (a) $J_2=0$ and (b) $J_2=0.15$. The lines are the linear fits of the data.} 
    \label{fig:del--mAF}
  \end{center}
\end{figure}

 In any case, the observation that the $S_{\bm q}$ peaks appear at mutually distinct positions at $J_2=0.3$ and 0.5 strongly suggests that the phases at $J_2=0.3$ and at $0.5$ are indeed different, the gapped I and II phases, and there is a phase transition between them.

 In Fig. \ref{fig:del00magnet-order}(a), we show the AF order parameter, the squared sublattice magnetization $m_{AF}^2$ associated with the standard two-sublattice order (the $\Gamma$-point order), plotted versus $1/\sqrt{N}$ for various values of $J_2$. It is defined by 
\begin{align}
m_{AF}^2 &=\frac{1}{2}\frac{1}{\frac{N}{4}(\frac{N}{4}+1)}
\left[\sum_{\alpha=A,B} \Braket{ \left( \sum_{i\in\alpha}{\bm S}_i
    \right)^2 } \right]_J , \nonumber \\
&=\frac{8}{N(N+4)} \left[\sum_{\alpha}\sum_{i,j\in \alpha}
\Braket{{\bm S}_i \cdot {\bm S}_j} \right]_J ,
\label{eq:m_AF}
\end{align}
where $\alpha=A,B$ denotes the two triangular sublattices of the original honeycomb lattice, and the sum over $i,j\in \alpha$ is taken over all sites $i,j$ belonging to the sublattice $\alpha$. When the system retains a relevant magnetic long-range order $m_{\infty}$, the size dependence proportional to  $1/\sqrt{N}$ is expected from the spin-wave analysis, {\it i.e.\/},
\begin{align}
m^2 = m_{\infty}^2 + \frac{c_1}{\sqrt{N}} ,
\label{eq:fit-sqrt}
\end{align}
where $c_1$ is a constant. As can be seen from the figure, a linear extrapolation of our finite-$N$  data indicates that the standard AF LRO vanishes at around $J_2=0.25$, consistently with the phase diagram shown in Fig. \ref{fig:del00phase}. \cite{Lauchli}

 We also compute the magnetic order parameter associated with the stripe order (the M-point order) $m_{str}^2$, defined by
\begin{align}
m_{str}^2 =\frac{8}{3N(N+4)}
\left[\sum_{\nu}\sum_{\alpha_\nu}\sum_{i,j\in \alpha_\nu} 
\langle {\bm S}_i \cdot {\bm S}_j\rangle \right]_J ,
\end{align}
where $\nu=1,2,$ and 3 refer to the three distinct types of stripe order associated with the threefold rotational symmetry of the lattice. The result is shown in Fig. \ref{fig:del00magnet-order}(b) as a function of $1/\sqrt{N}$ for $J_2=0.3, 0.4$, and 0.5, where we restrict the sizes only to $N=8$, 24, and 32, which retain the threefold lattice rotational symmetry being commensurate with the stripe order. As can be seen from the figure, $m_{str}^2$ is always extrapolated to negative values for $N\rightarrow \infty$, which indicates that the stripe order remains a SRO in both the gapped I and II phases.

 We also compute the spin-gap energy, {\it i.e.\/}, the energy difference $\Delta E$ between the ground state and the triplet first-excited state. It turns out that $\Delta E$ is extrapolated to zero for the AF state of $J_2\lesssim 0.25$, while to a positive nonzero value for the gapped I and II states of $J_2\gtrsim 0.25$. Some of the data will be shown in sect. 4 below (see Fig. \ref{fig:del--gap}).

\section{Random model: Phase diagram and ground-state properties}
\label{sec:del--}

\subsection{Phase diagram}

In this section, we present our numerical results on the random $(\Delta\not=0)$ model in the region of $0\leq J_2\leq 0.5$. We first show our main result in Fig. \ref{fig:phasediagram-random}, the ground-state phase diagram of the bond-random $s=1/2$ $J_1-J_2$ Heisenberg model on the honeycomb lattice in the frustration ($J_2$) versus the randomness ($\Delta$) plane. The  $\Delta=0$ line corresponds to the phase diagram of the regular model shown in Fig. \ref{fig:del00phase} in the previous section. In fact, when the randomness $\Delta$ is sufficiently weak, the phase diagram turns out to be qualitatively unchanged from that of the regular model.

 In Fig. \ref{fig:phasediagram-random}, four distinct phases are identified. Three of them, {\it i.e.\/}, the AF phase, and the gapped phases I and II, have already been identified in the regular model, while, when the strength of the randomness exceeds a critical value $\Delta_c(J_2)$, the fourth phase, the random-singlet phase, appears. In fact, the random-singlet phase turns out to be stabilized for a wide range of parameter space for moderate or strong randomness. As will be shown shortly, this phase is basically of the same nature as the one recently identified in the $s=1/2$ random triangular and kagome Heisenberg models. \cite{Watanabe,Kawamura,Shimokawa}

Below, we show our numerical data for various observables including the order parameters, the spin-gap energy, and the static and dynamical structure factors for the $J_2$-regions of (i) $0\leq J_2\leq 0.25$ and (ii) $0.25\leq J_2\leq 0.5$, separately.

\begin{figure}9a
  \begin{center}
    \includegraphics[width=7.5cm]{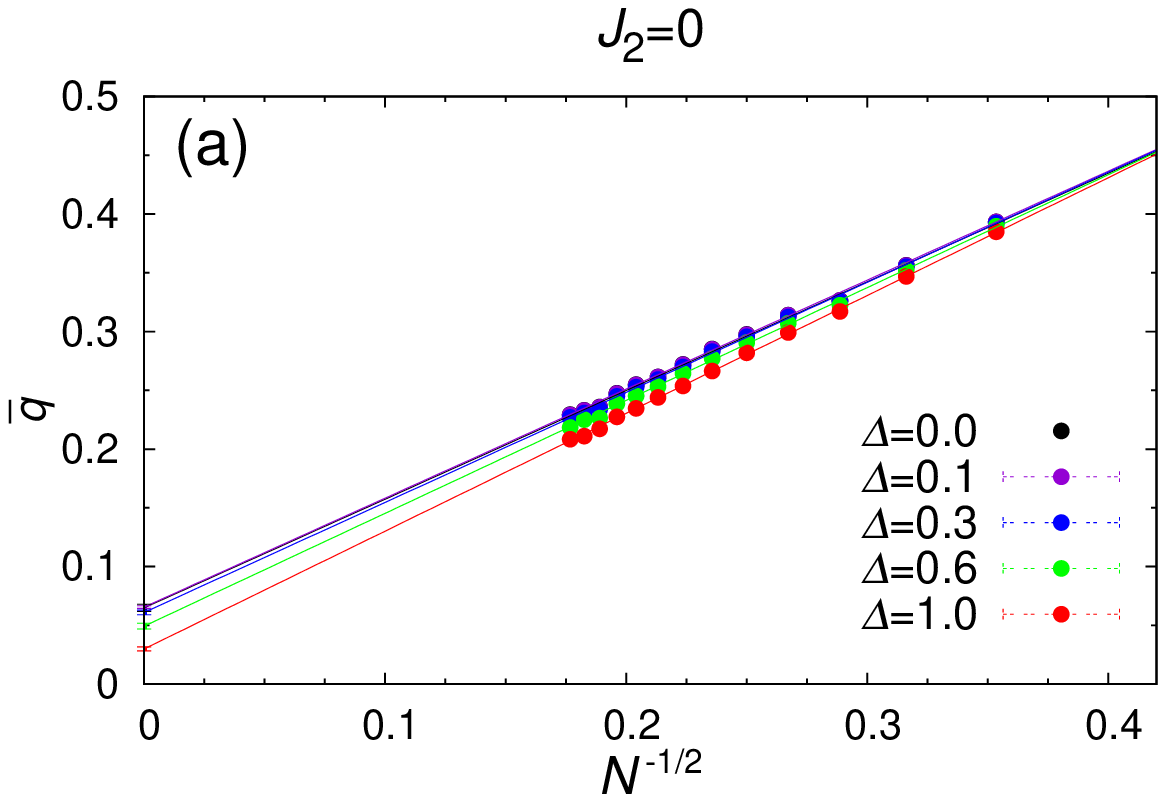}
    \includegraphics[width=7.5cm]{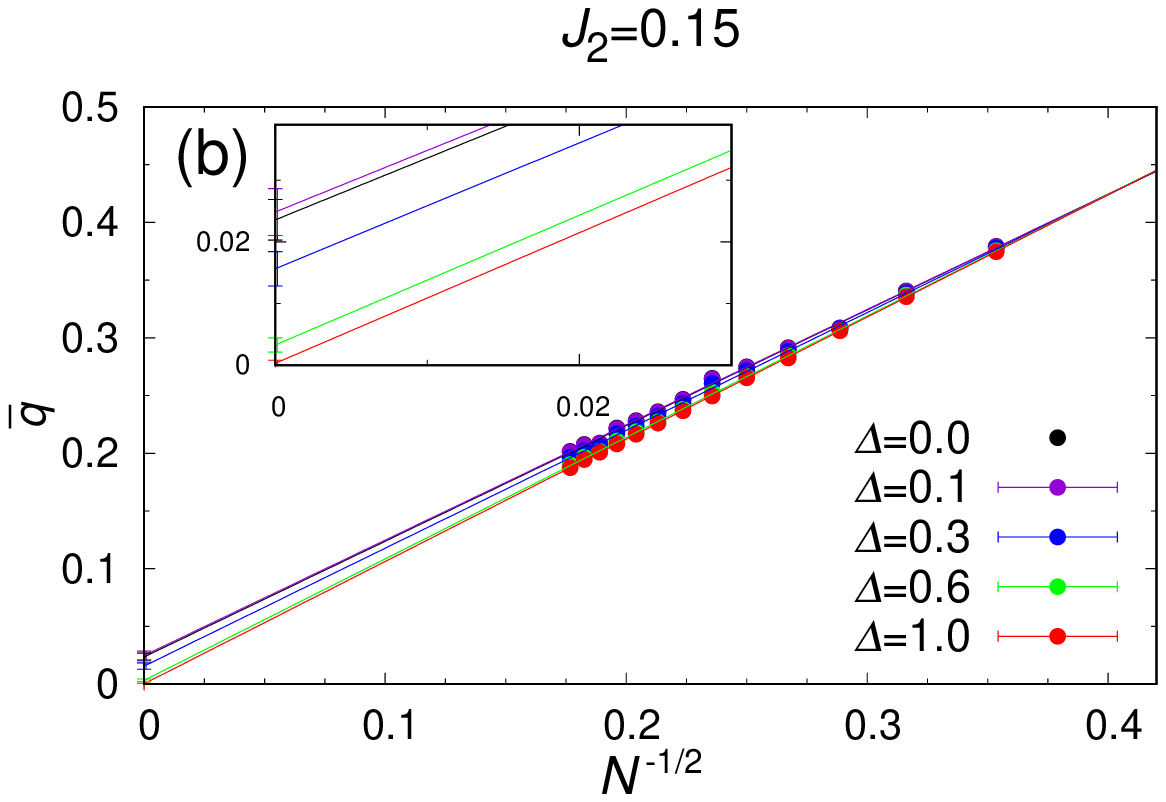}
    \caption{Spin freezing parameter $\bar{q}$ of the random model plotted versus  $1/\sqrt{N}$ for various values of $\Delta$, at (a) $J_2=0$ and (b) $J_2=0.15$. The lines are linear fits of the data. The inset of (b) is a magnified view of the large-$N$ region.}
    \label{fig:del--qb1}
  \end{center}
\end{figure}
\begin{figure*}[t]
  \begin{center}
    \begin{tabular}{c}
      \begin{minipage}{0.33\hsize}
        \begin{center}
          \includegraphics[clip,width=0.99\hsize]{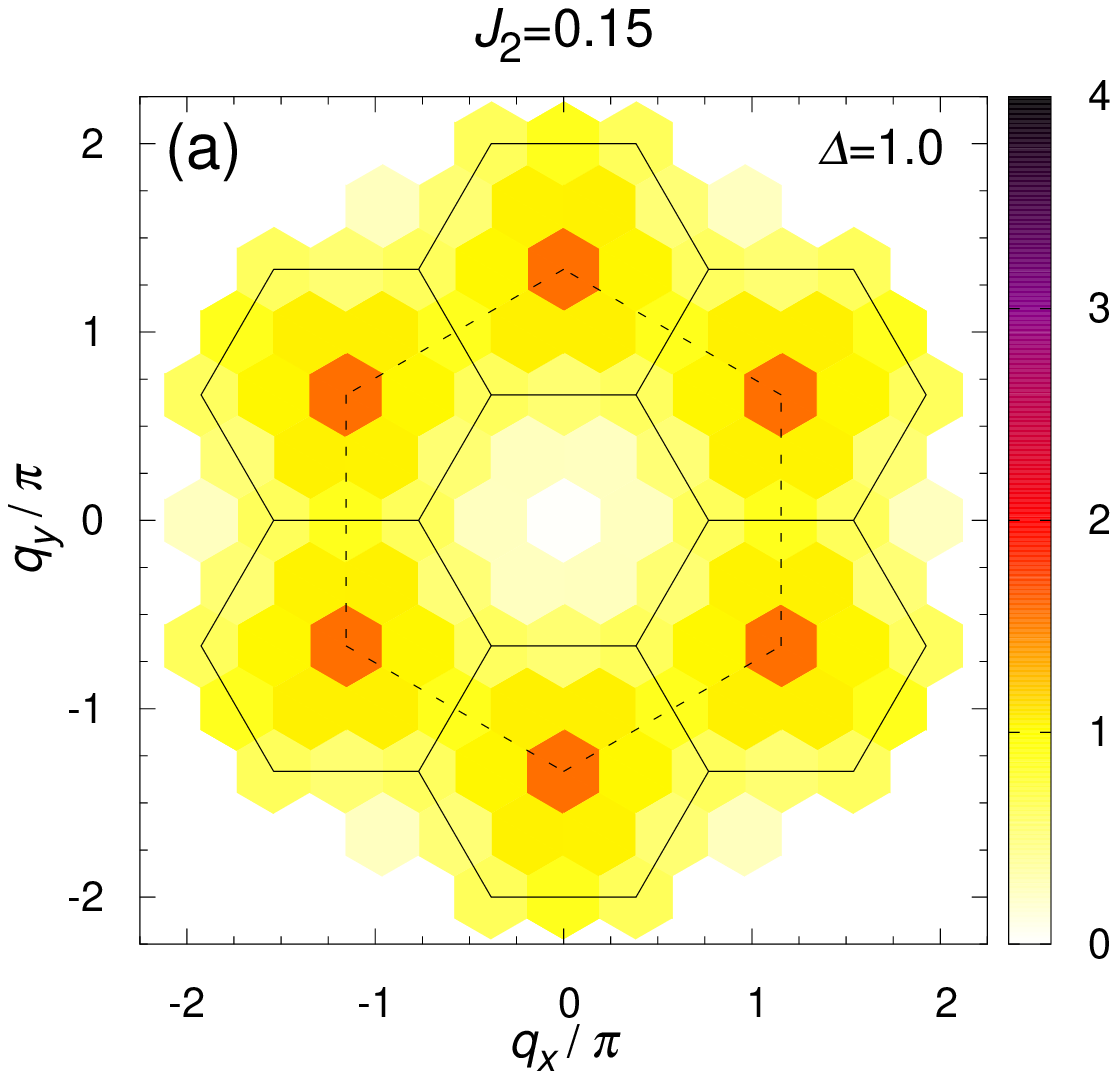}
       \end{center}
      \end{minipage}
      \begin{minipage}{0.33\hsize}
        \begin{center}
          \includegraphics[clip,width=0.99\hsize]{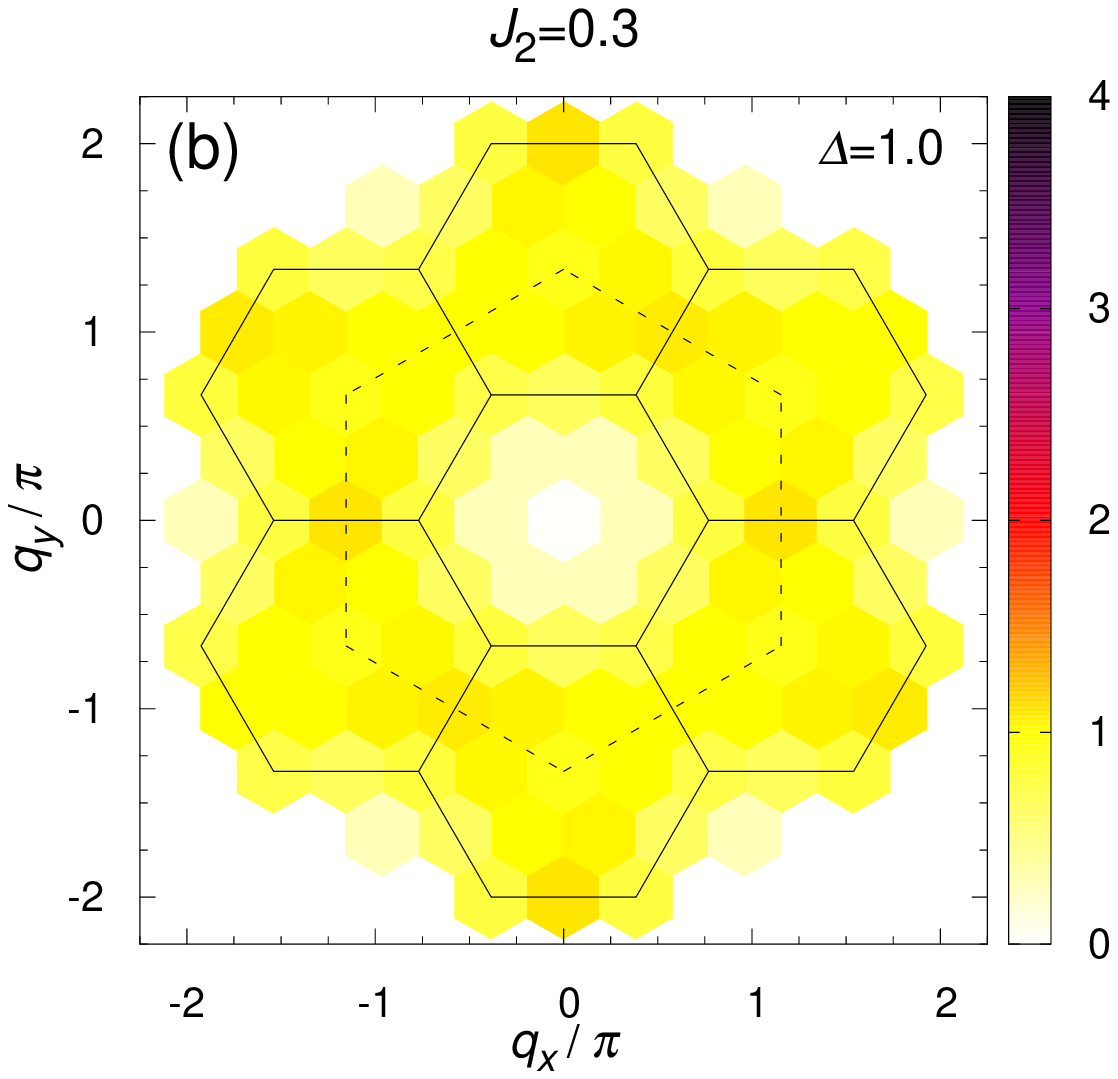}
        \end{center}
      \end{minipage}
      \begin{minipage}{0.33\hsize}
        \begin{center}
          \includegraphics[clip,width=0.99\hsize]{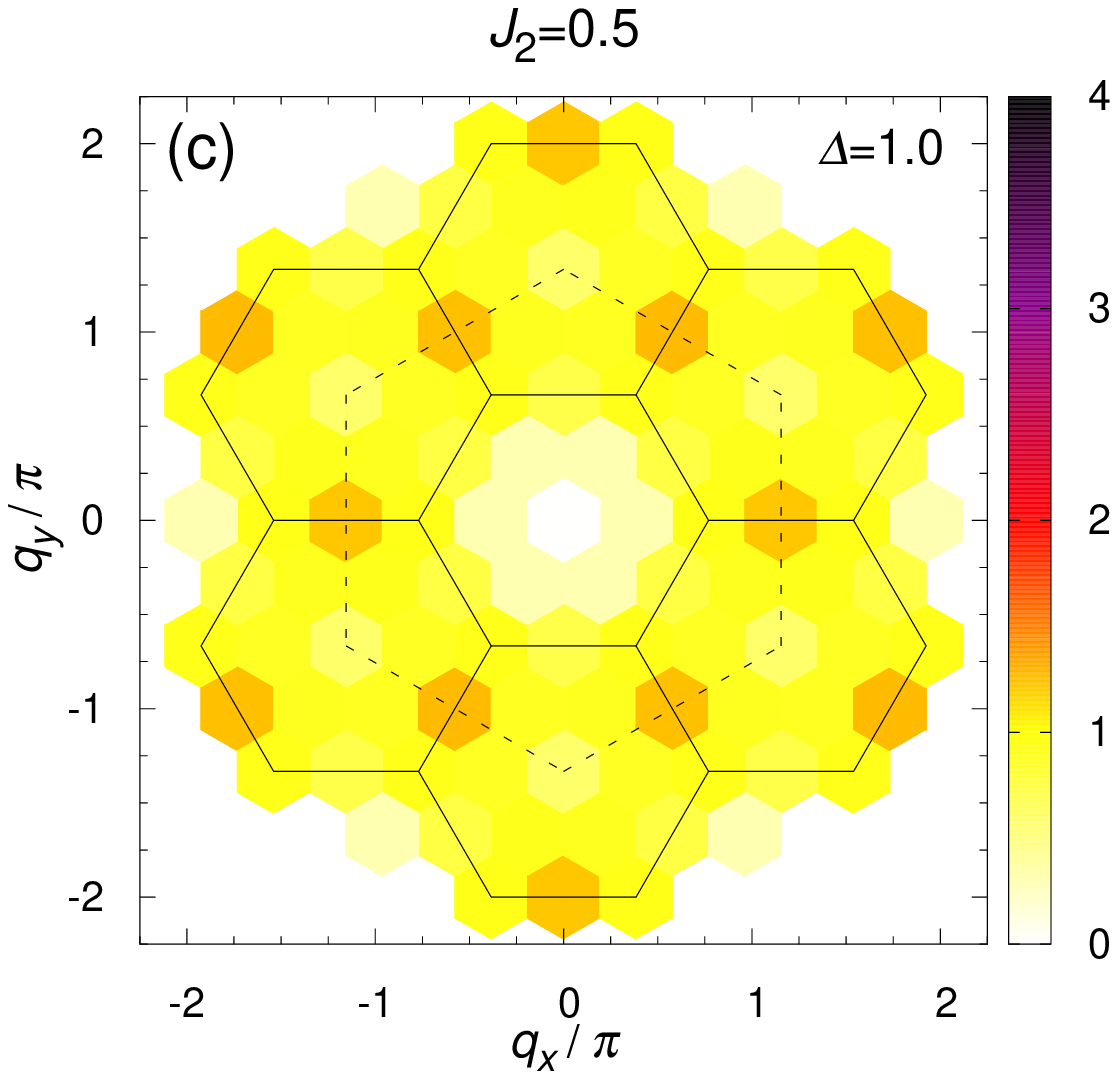}
        \end{center}
      \end{minipage}
    \end{tabular}
    \caption{Intensity plots of the static spin structure factor $S_{\bm q}$ of the random model at (a) $J_2=0.15$, (b) $J_2=0.3$, and (c) $J_2=0.5$. The lattice size is $N=24$. The solid line shows the boundary of the first Brillouin zone of the triangular lattice composing a sublattice of the original honeycomb lattice. The dashed line shows the boundary of the extended  Brillouin zone.}
    \label{fig:SSF-random}
  \end{center}
\end{figure*}

\subsection{Region $0\leq J_2 \leq 0.25$}

First, we investigate the $0\leq J_2\leq 0.25$  region where the standard AF order is stable in the regular model of $\Delta=0$. In Fig. \ref{fig:del--mAF}, we show the squared sublattice magnetization $m_{AF}^2$ plotted versus $1/\sqrt{N}$ for the cases of (a) $J_2=0$ and (b) $J_2=0.15$, for various values of the randomness $\Delta$ spanning from the regular case of $\Delta=0$ to the maximal randomness case of $\Delta=1$.

 For $J_2=0$, {\it i.e.\/}, for the nearest-neighbor model without frustration, Fig. \ref{fig:del--mAF}(a) shows that $m_\infty^2$ is extrapolated to positive values for any $\Delta$, indicating the AF LRO stabilized up to the maximal randomness. This behavior of the random honeycomb-lattice nearest-neighbor model differs from that of the random triangular-lattice nearest-neighbor model where the AF order becomes unstable at a finite amount of randomness of $\Delta_c\simeq 0.5$, giving way to the random-singlet state for stronger randomness $\Delta>\Delta_c$.\cite{Watanabe} 

 By contrast, as can be seen from Fig. \ref{fig:del--mAF}(b), for $J_2=0.15$, there exists a finite critical randomness of $\Delta_c\simeq 0.6$ beyond which the AF LRO vanishes. This observation, together with the previous data on the random triangular model, \cite{Shimokawa} demonstrates that a certain amount of frustration is necessary to destabilize the AF LRO by introducing the randomness. We have made a similar analysis for other values of $J_2\leq 0.25$, and draw the AF-random-singlet phase boundary (marked by red points) in the phase diagram of Fig. \ref{fig:phasediagram-random}.

 In order to investigate the possible appearance of other types of magnetic order, we compute the spin freezing parameter $\bar q$ defined by
\begin{align}
\bar{q}= \frac{1}{N}
\sqrt{\left[\sum_{i,j}\Braket{{\bm S}_i\cdot{\bm S}_j}^2\right]_J}.
\label{eq:q_bar}
\end{align}
This quantity can detect any type of static spin order, even including the random one such as the spin-glass order.  The computed $1/\sqrt{N}$-dependence of $\bar q$ is shown in Fig. \ref{fig:del--qb1} for the cases of (a) $J_2=0$ and (b) $J_2=0.15$. The inset exhibits a magnification of the larger $N$ region. Note that, when the system possesses the AF LRO, $\bar q$ also becomes nonzero. The interest here is whether $\bar q$ could be nonzero in the parameter region without the AF LRO. As can be seen from these figures,  whether the extrapolated $\bar q$ is positive or negative (zero) well correlates with the behavior of $m_{AF}^2$ shown in Fig. \ref{fig:del--mAF}, indicating that no magnetically ordered state other than the standard AF order appears in the parameter range studied. In particular, no indication of the spin-glass order stabilized by the introduced randomness is obtained. This means that the state observed at $J_2=0.15$ for a stronger randomness of $\Delta \gtrsim 0.6$ is a nonmagnetic state without any static spin order.

 We also compute the mean spin-gap energy $\Delta E=[\Delta E]_J$ of the random model (data not shown here), to find that $\Delta E$ is always extrapolated to vanishing values within the error bar, indicating that the nonmagnetic state found in the $\Delta\gtrsim 0.6$ region for $J_2=0.15$ is gapless, distinct from the gapped I and II phases.

 We also compute the static and dynamical spin structure factors. In Fig. \ref{fig:SSF-random}, we show the static spin structure factors for the maximally random case of $\Delta=1$ for various $J_2$-values. For the case of $J_2=0.15$, as can be seen from Fig. \ref{fig:SSF-random}(a), the peaks appear at the same $q$-points as the corresponding regular one, {\it i.e.\/}, at the $\Gamma$ point, although they are much broadened as compared with the regular case, reflecting the SRO nature of the associated AF order.

 The dynamical spin structure factor  $S_{\bm q}(\omega)$ is defined by
\begin{align}
S_{\bm q}(\omega)&=\int_{-\infty}^\infty
\left[ \Braket{\left(S_{\bm q}^z(t)\right)^\dag
  S_{\bm q}^z(0)} \right]_Je^{-i\omega t}dt \nonumber \\
&=-\lim_{\eta\to 0}\left[\frac{1}{\pi}{\rm Im}\Braket{
    (S_{\bm q}^z)^\dag\frac{1}{\omega + E_0 + i\eta -\mathcal{H}}
    S_{\bm q}^z}\right]_J,
\label{eq:DSF}
\end{align}
where $E_0$ is the ground-state energy, and $\eta$ is a phenomenological damping factor taking a sufficiently small positive value. We employ the continued fraction method to compute $S_{\bm q}(\omega)$, \cite{Gagliano} putting $\eta=0.02$.

\begin{figure}
  \begin{center}
    \includegraphics[width=7.5cm]{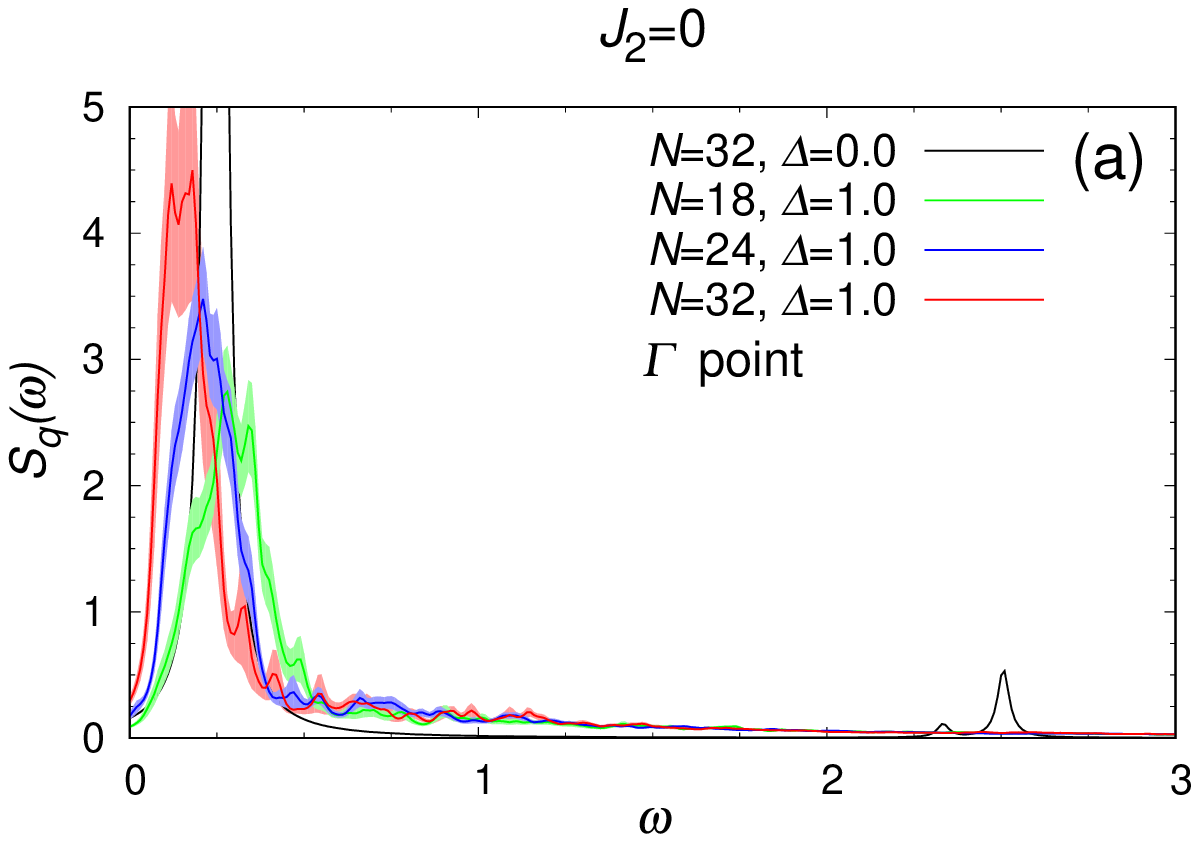}
    \includegraphics[width=7.5cm]{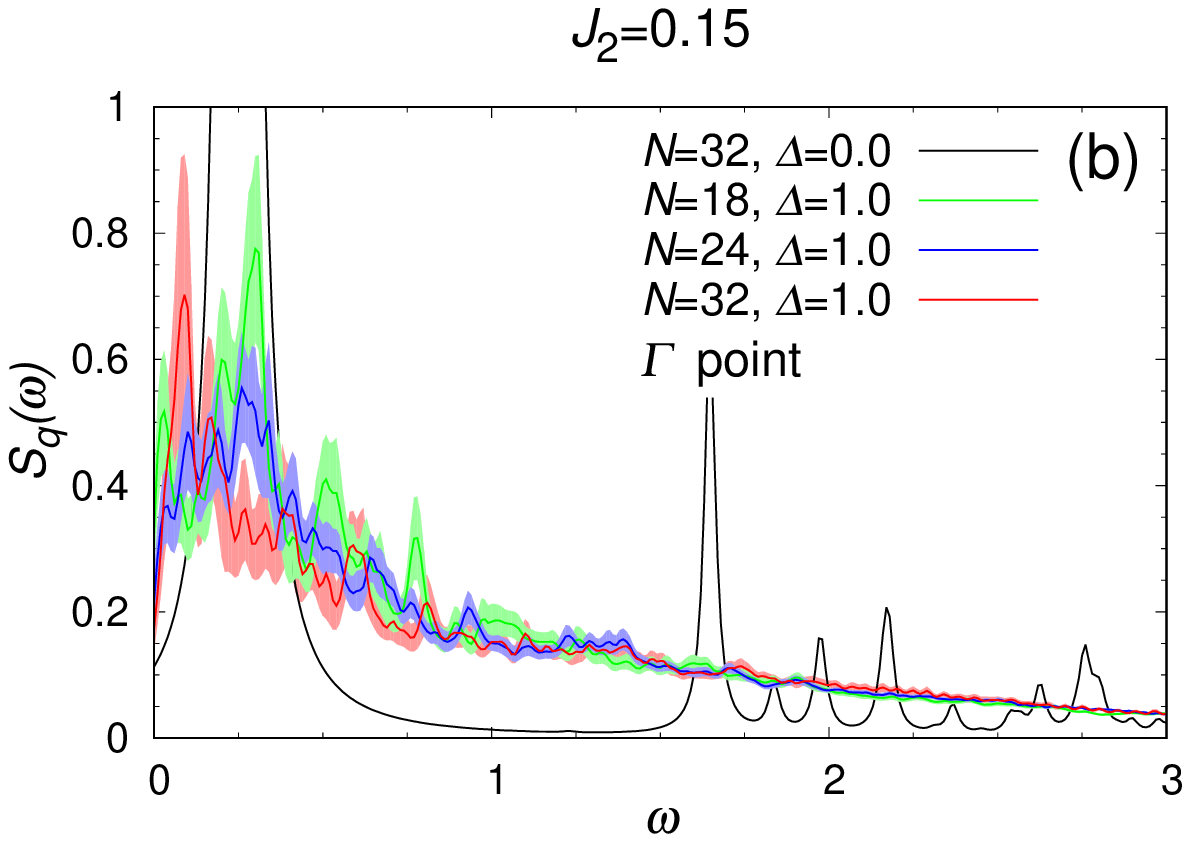}
    \caption{The dynamical spin structure factors $S_{\bm q}(\omega)$ of the random model of $\Delta=1$ computed at the  $\Gamma$ point at (a) $J_2=0$, and at (b) $J_2=0.15$, as compared with those of the regular model of $\Delta=0$. The lattice size is $N=18$, 24, and 32. Error bars are represented by the width of the data curves.} 
    \label{fig:DSF1}
  \end{center}
\end{figure}

 The $\omega$-dependence of $S_{\bm q}(\omega)$ computed at the $\Gamma$ point and at the maximal randomness of $\Delta=1$ is shown in Fig. \ref{fig:DSF1} for the cases of (a) $J_2=0$ and (b) $J_2=0.15$. In both Figs. \ref{fig:DSF1}(a) and \ref{fig:DSF1}(b), a sharp peak is observed in the small-$\omega$ region. In the regular case of $\Delta=0$, this peak is a finite-size counterpart of the delta-function peak expected at $\omega=0$ in the thermodynamic limit. Close inspection reveals that the small-$\omega$ peak in the maximally random case of $\Delta=1$ exhibits mutually different behaviors between (a) $J_2=0$ and (b) $J_2=0.15$. In Fig. \ref{fig:DSF1}(a), the peak tends to grow in height, while its position tends to move toward $\omega=0$  as the size $N$ is increased. By contrast, in Fig. \ref{fig:DSF1}(b), the peak height tends to saturate even when the system size $N$ is increased. This difference corresponds to the observation that in the maximally random case, the AF LRO is present for $J_2=0$, but is absent for $J_2=0.15$. Another interesting observation is that  $S_{\bm q}(\omega)$ at $J_2=0.15$ and $\Delta=1$, corresponding to the nonmagnetic state, exhibits a very broad background component and a tail extending to larger values of $\omega$, in addition to a relatively sharp peak at a smaller $\omega$. Such a broad feature is a characteristic of the random-singlet state as previously studied in detail for the cases of the bond-random triangular and kagome models. \cite{Shimokawa} In fact, our data of $S_{\bm q}(\omega)$ resembles the corresponding one of the random triangular model. This is the second indication that the nonmagnetic state observed for the $\Delta\gtrsim 0.6$ region at $J_2=0.15$ is indeed a random-singlet state.

\subsection{Region $0.25\leq J_2 \leq 0.5$}

\begin{figure}
  \begin{center}
    \includegraphics[width=7.5cm]{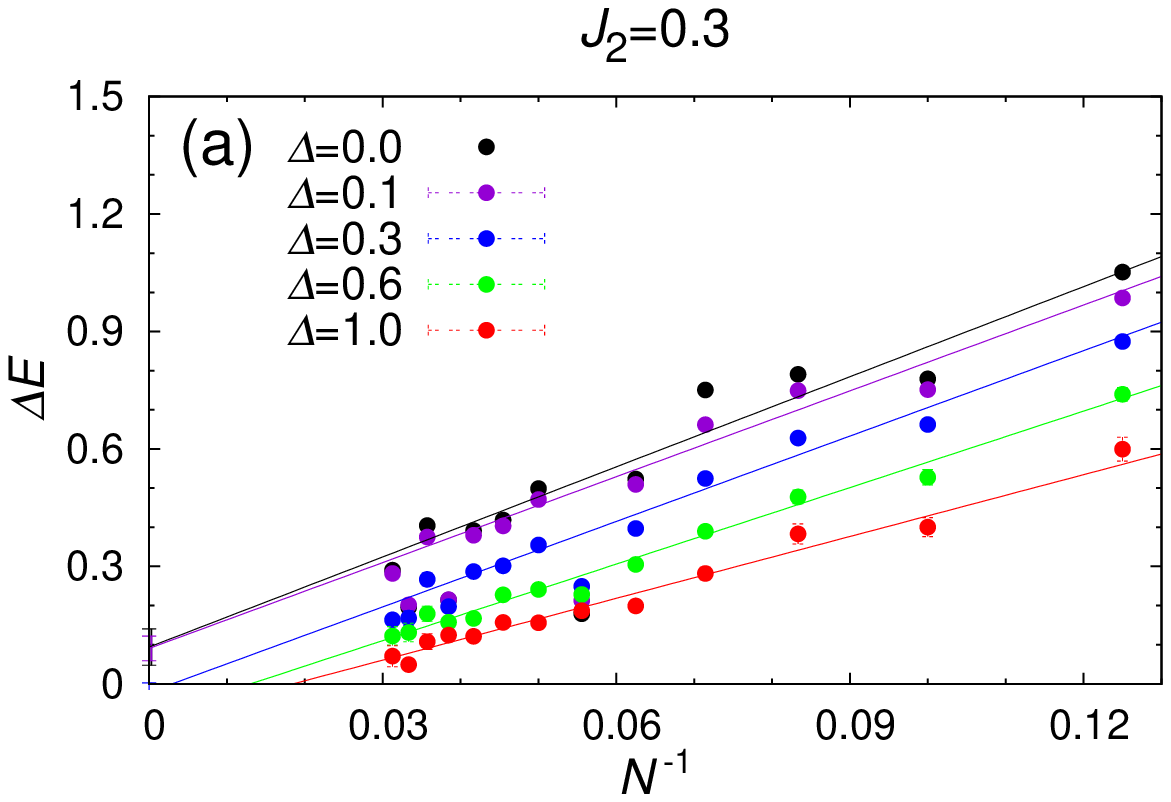}\par
    \includegraphics[width=7.5cm]{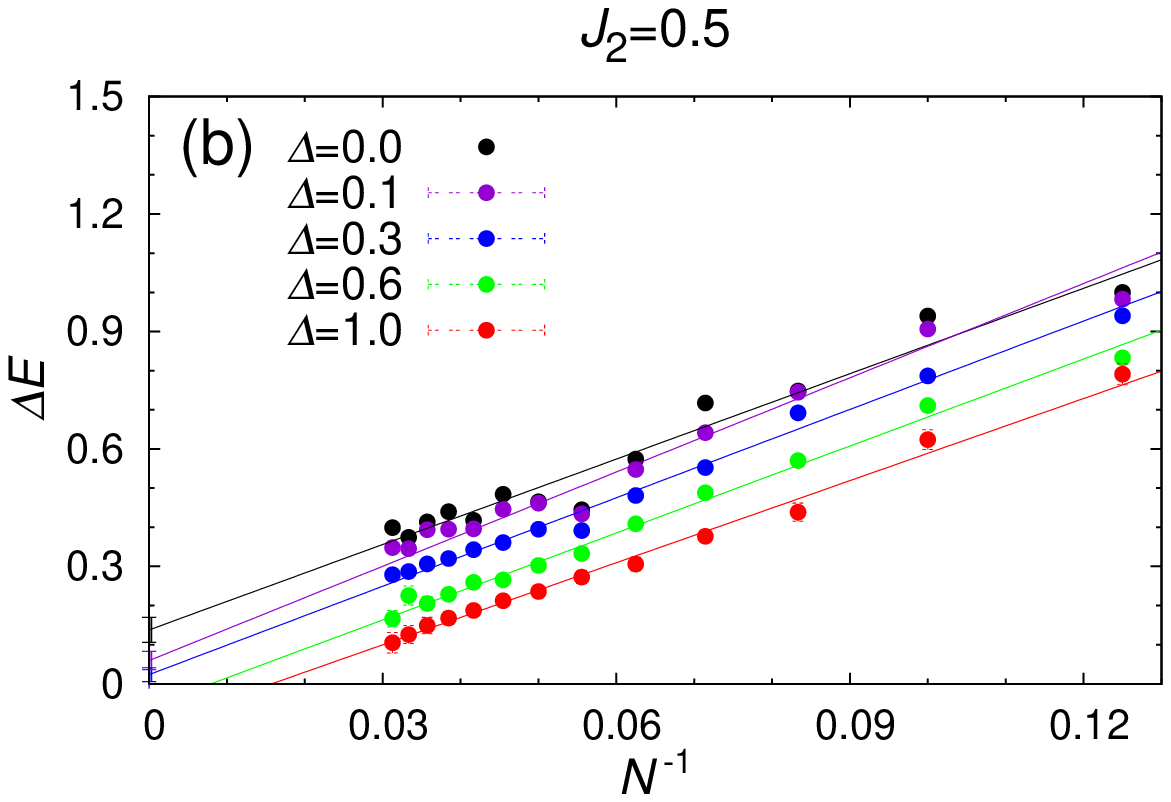}\par
    \caption{Mean spin-gap energy $\Delta E$ for various values of $\Delta$ plotted versus $1/N$, at (a) $J_2=0.3$ and (b) $J_2=0.5$. The lines are linear fits of the data. In the fit of (a), the data points of $N=18$, 26, and 30 are excluded. 
      }
    \label{fig:del--gap}
  \end{center}
\end{figure}
\begin{figure}
  \begin{center}
    \includegraphics[width=7.5cm]{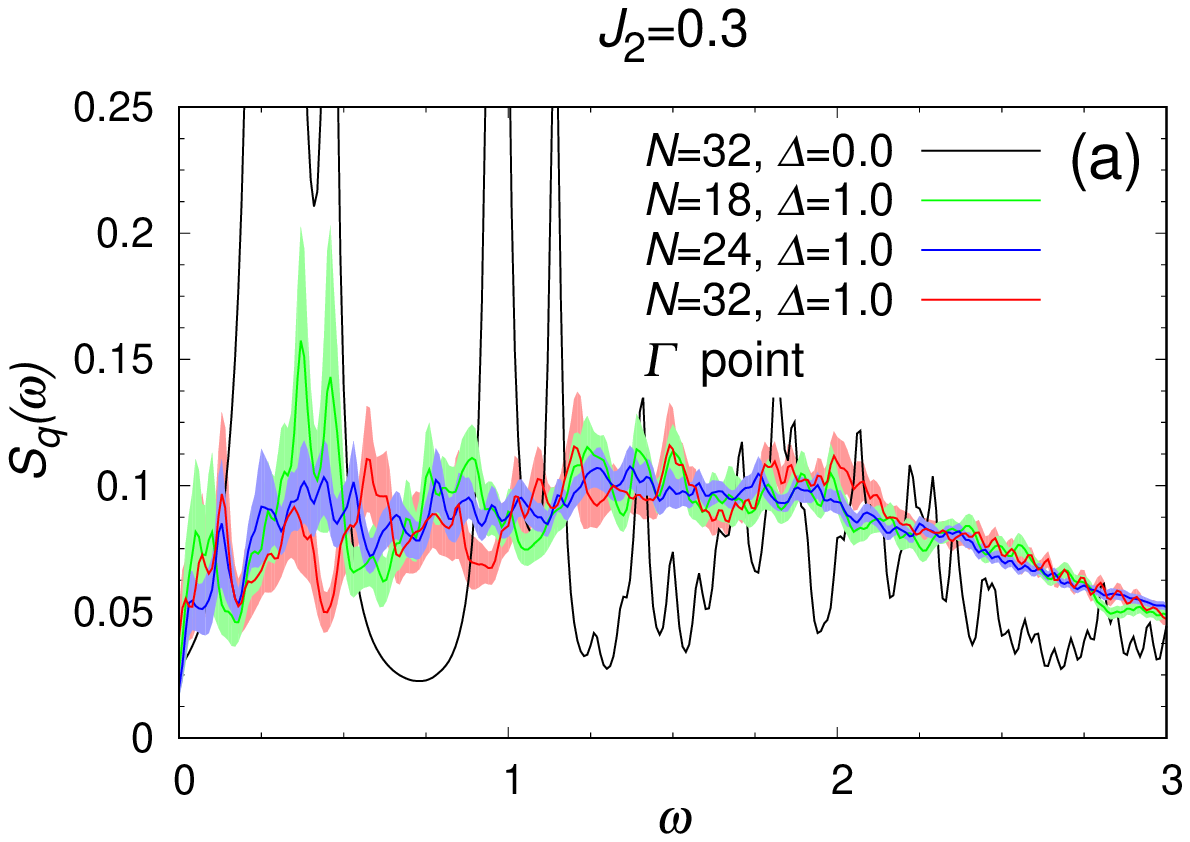}
    \includegraphics[width=7.5cm]{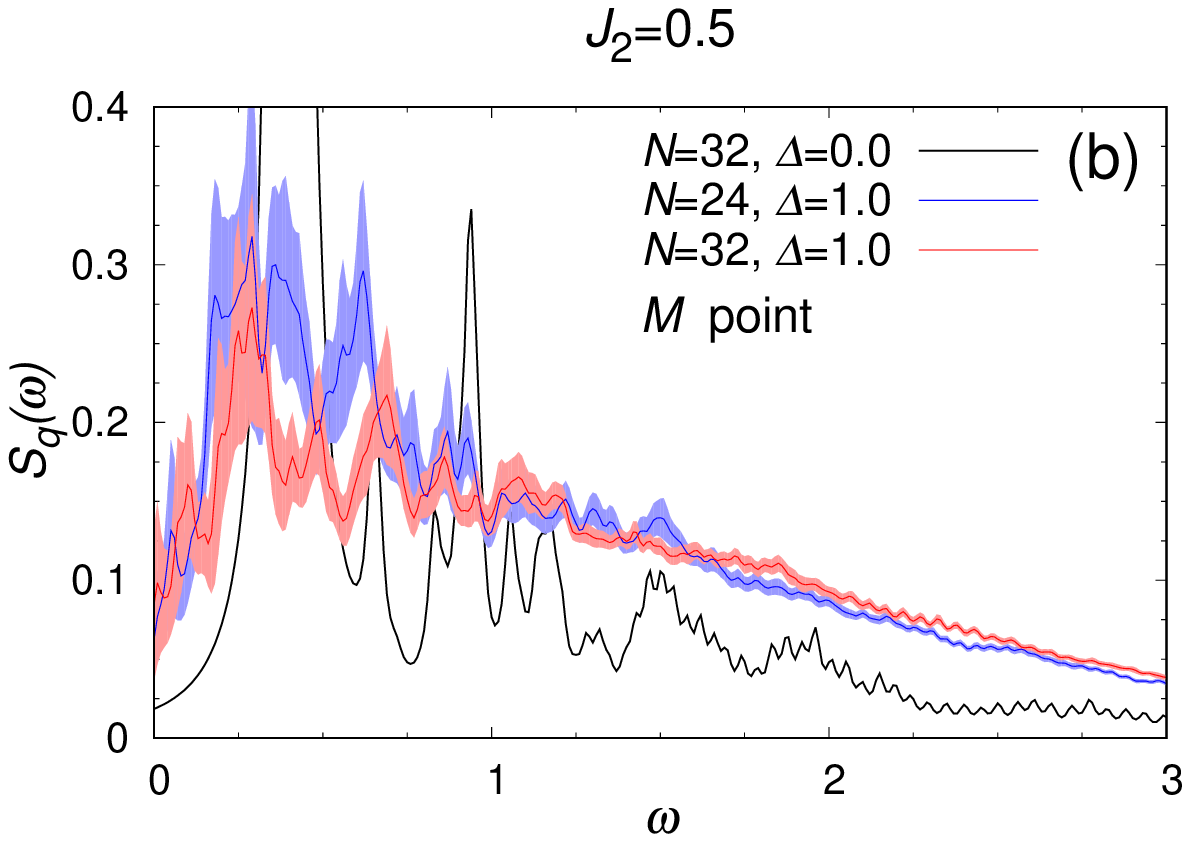}
   \caption{Dynamical spin structure factors $S_{\bm q}(\omega)$ of the random model of $\Delta=1$ computed at (a) the $\Gamma$ point at $J_2=0.3$ and (b) the $M$ point at $J_2=0.5$, as compared with those of the regular model of $\Delta=0$. The lattice size is $N=18$, 24, and 32. Error bars are represented by the width of the data curves.}
    \label{fig:DSF2}
  \end{center}
\end{figure}

\begin{figure}
  \begin{center}
    \includegraphics[width=7.5cm]{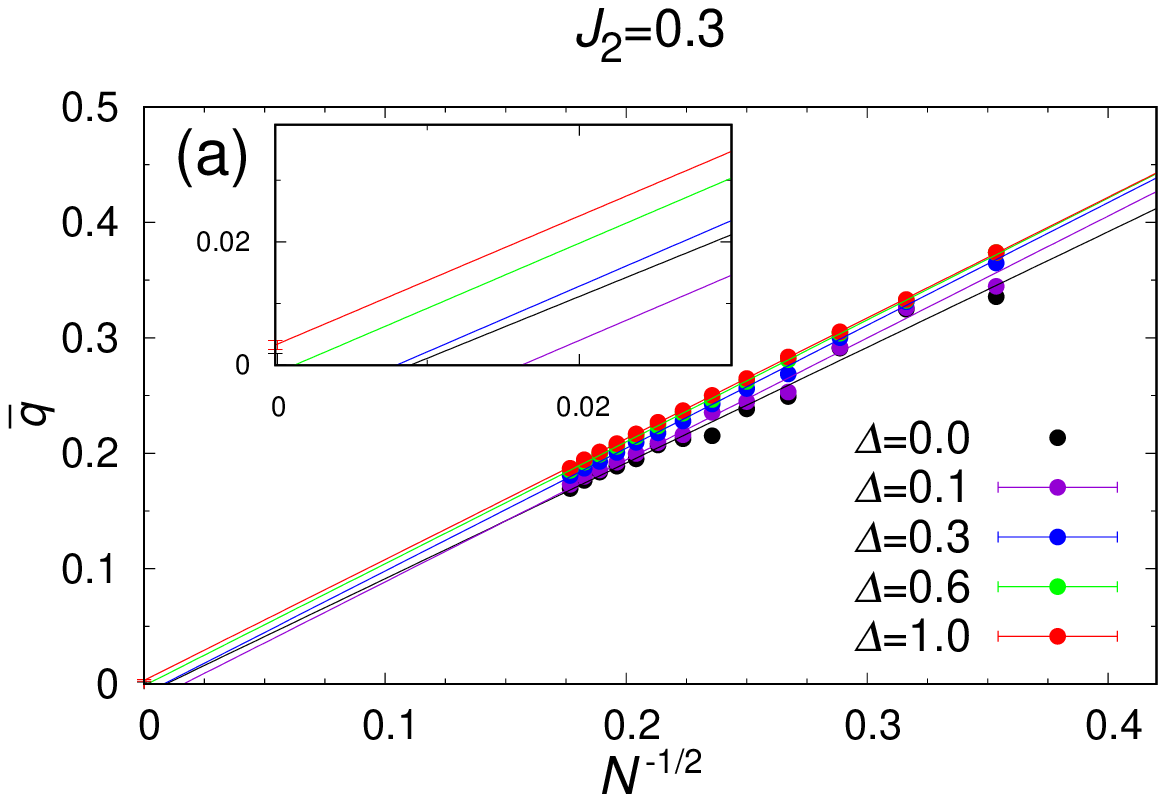}\par
    \includegraphics[width=7.5cm]{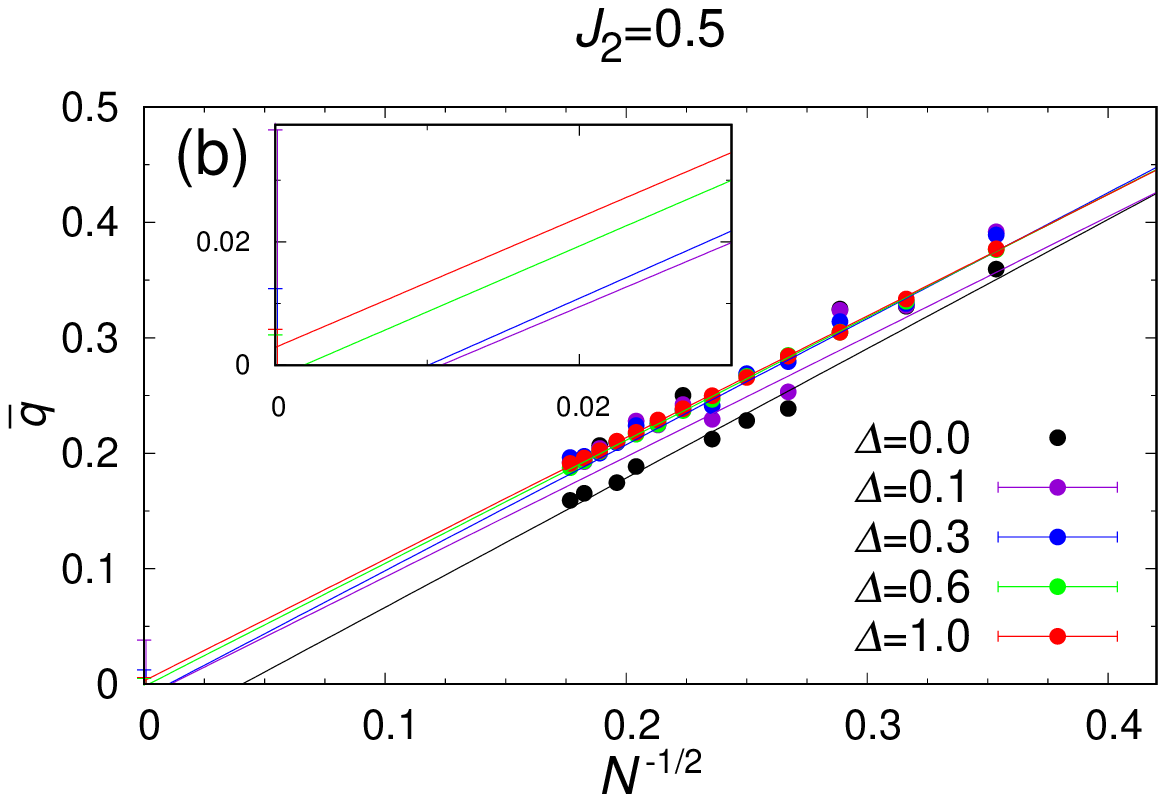}\par
\caption{Spin freezing parameter $\bar{q}$ plotted versus $1/\sqrt{N}$ for various values of $\Delta$, at (a) $J_2=0.3$ and (b) $J_2=0.5$. Lines are linear fits of the data. Insets are magnified views of the large-$N$ region. In the fit of (b), only data points observing the threefold lattice rotational symmetry are used.}
    \label{fig:del--qb2}
  \end{center}
\end{figure}

 Next, we move to the larger-$J_2$ region of $0.25\leq J_2\leq 0.5$, which corresponds to the gapped I and II phases of the regular model of $\Delta=0$. In order to obtain information of the possible magnetic  LRO, we compute the size dependence of the freezing parameter $\bar q$, and the result is shown in Fig. \ref{fig:del--qb2} for (a) $J_2=0.3$ and (b) $J_2=0.5$, each corresponding to the gapped I and gapped II phases, respectively. As can be seen from the figures, $\bar q$ is extrapolated to zero within the error bar for any value of $\Delta$, indicating that the ground state in this region is always nonmagnetic. (Precisely speaking, for the case of $\Delta=1$, the fit using all the data points yields a slightly positive $\bar{q}$, while the fit using only larger-$N$ data points of $N\ge24$ yields a vanishing $\bar{q}$ within one $\sigma$.)

In Fig. \ref{fig:del--gap}, we show the size dependence of the spin-gap energy $\Delta E$ for (a) $J_2=0.3$ and (b) $J_2=0.5$. Interestingly, the extrapolated $\Delta E$ is zero, {\it i.e.\/}, the system is gapless for larger $\Delta>\Delta_c$, while it becomes nonzero, {\it i.e.\/}, gapped for smaller $\Delta<\Delta_c$. (In our data fit of Fig. \ref{fig:del--gap}(b), the data of $N=18$, 26, and 30 for $J_2=0.3$ are excluded since they largely deviate from other data.) The values of $\Delta_c$ are estimated to be $\Delta_c\simeq 0.2$ for $J_2=0.3$ and $\Delta_c\simeq 0.4$ for $J_2=0.5$. The gapped states for smaller $\Delta$ correspond to the gapped I and II states discussed in the previous section. The change observed between the gapless and gapped behaviors on increasing $\Delta$ suggests the occurrence of a randomness-induced phase transition. The gapless nonmagnetic phase stabilized at $\Delta > \Delta_c$ is likely to be the random-singlet state. 

 Further information can be obtained from the static and dynamical spin structure factors. Figure \ref{fig:SSF-random} has shown the static spin structure factor for the maximally random case of $\Delta=1$ for various $J_2$-values. For the case of $J_2=0.5$, as can be seen from Fig. \ref{fig:SSF-random}(c), the peak appears at the same $q$-points as the corresponding regular case, {\it i.e.\/}, at the M points. Meanwhile, for the case of $J_2=0.3$ shown in Fig. \ref{fig:SSF-random}(b), the peak structure itself is hardly discernible.

 In Fig. \ref{fig:DSF2}, we show the $\omega$-dependence of the corresponding dynamical spin structure factor $S_{\bm q}(\omega)$ for $\Delta=1$, {\it i.e.\/}, at (a) $J_2=0.3$ computed at the $\Gamma$ point, and at (b) $J_2=0.5$ computed at the M points. As can be seen from the figures, the computed $S_{\bm q}(\omega)$ exhibits much less peaky behavior as compared with the regular case, possessing a very broad component with a long tail extending to a larger $\omega$. This feature is a characteristic of the random-singlet state of the random triangular and kagome models studied in Refs. \cite{Kawamura} and \cite{Shimokawa}. (The data of Fig. \ref{fig:DSF2} appear to resemble the random kagome model more than the random triangular model.) Anyway, such resemblance also justifies our present identification of the randomness-induced gapless nonmagnetic state stabilized for larger $\Delta$ as the random-singlet state, as observed in the $0\leq \Delta \leq 0.25$ region.

 Collecting all the results above, we construct the ground-state phase diagram in the $J_2$-$\Delta$ plane as shown in Fig. \ref{fig:phasediagram-random}. The phase boundary between the AF state and the random-singlet state (red points in Fig. \ref{fig:phasediagram-random}) is determined from the behavior of  $m_{AF}^2$, while that between the gapped I and II phases and the random-singlet phase (blue points in Fig. \ref{fig:phasediagram-random}) is determined from the spin gap $\Delta E$. The phase boundary between the gapped I and II phases (green points in Fig. \ref{fig:phasediagram-random}) is determined from the peak location of $S_{\bm q}$.

\section{Random model: Finite-temperature properties}
\label{sec:finitemp}

 In this section, we investigate the finite-temperature properties of the random $J_1$-$J_2$ Heisenberg model on the honeycomb lattice, focusing on the specific heat and the uniform susceptibility. To compute the thermal average of physical quantities, we employ the Hams--de Raedt method. \cite{HamsRaedt} Following the previous sections, we present the results for the regions of (i) $0\leq J_2\leq 0.25$ and (ii) $0.25\leq J_2\leq 0.5$, separately.

\subsection{Region $0\leq J_2\leq 0.25$}

\begin{figure}
  \begin{center}
    \begin{tabular}{c}
      \begin{minipage}{0.5\hsize}
        \begin{center}
          \includegraphics[clip,width=0.99\hsize]{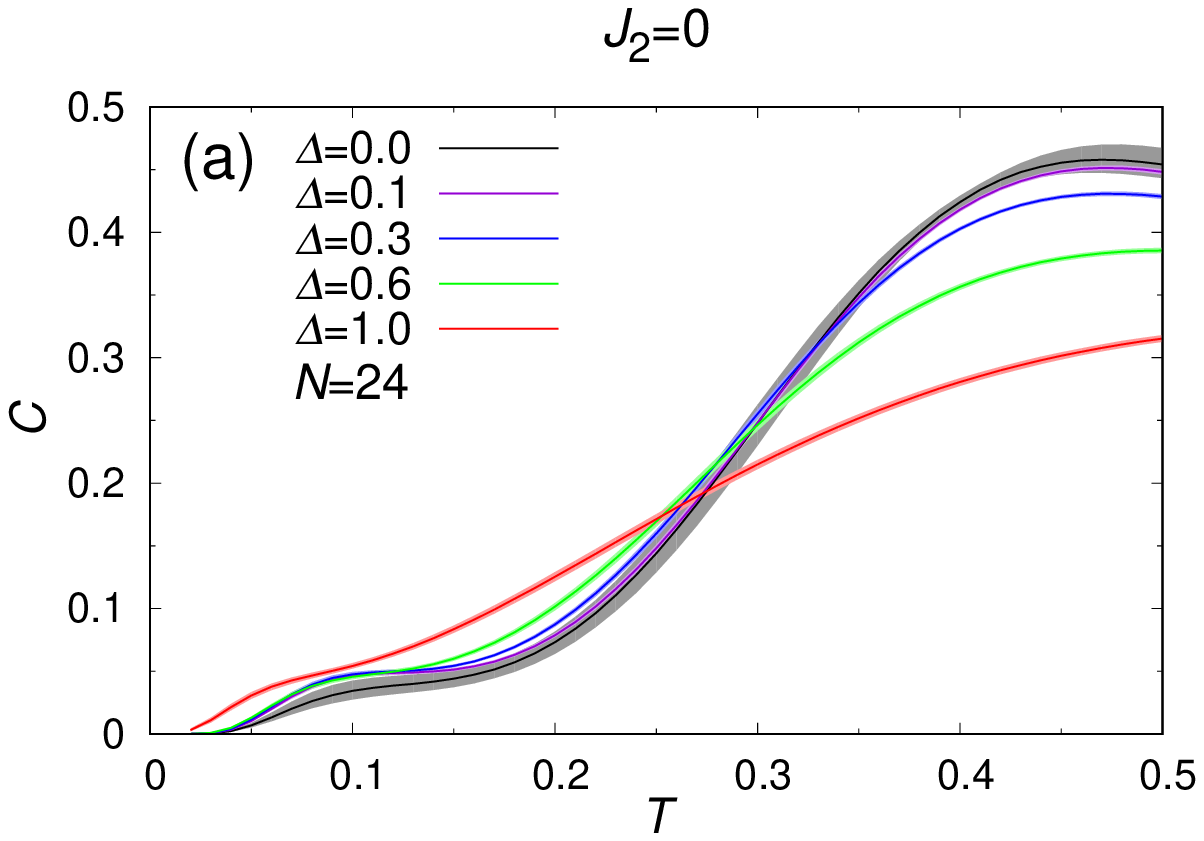}
        \end{center}
      \end{minipage}
      \begin{minipage}{0.5\hsize}
        \begin{center}
          \includegraphics[clip,width=0.99\hsize]{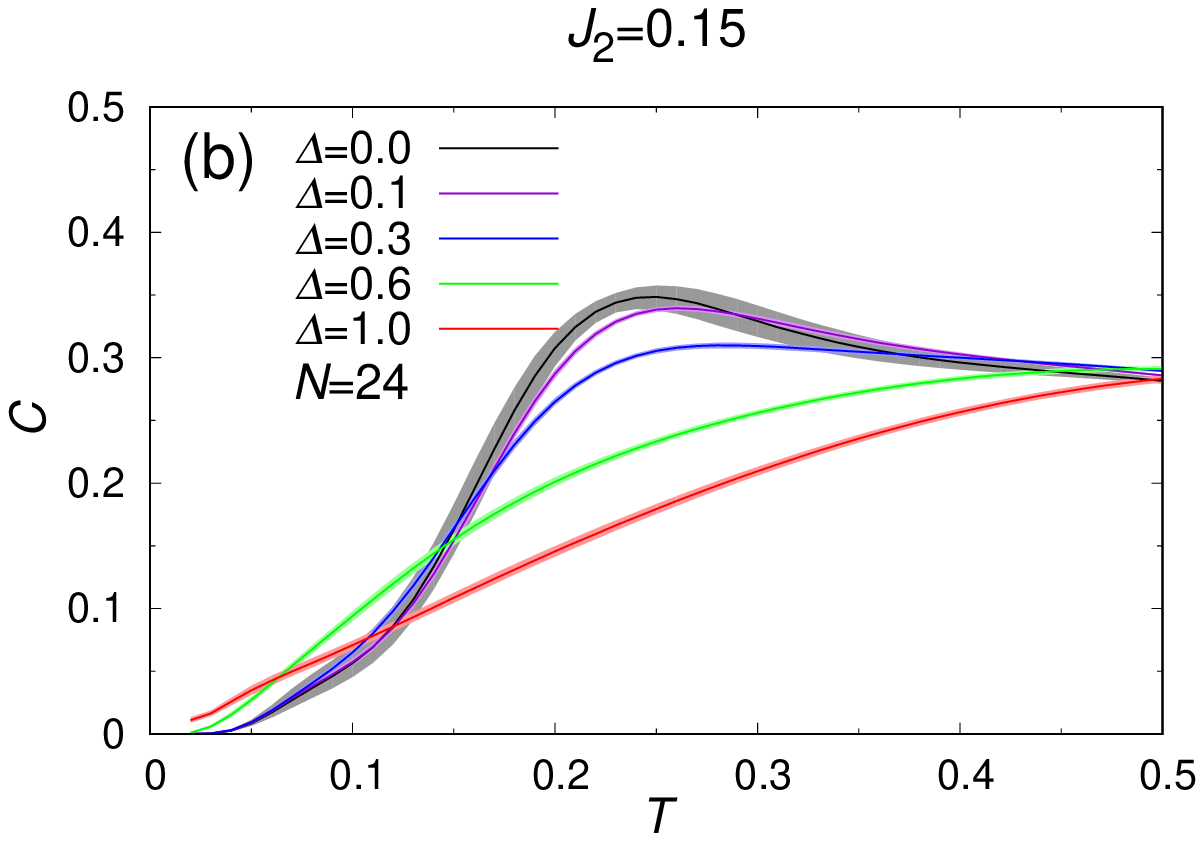}
        \end{center}
      \end{minipage}\\
      \begin{minipage}{0.5\hsize}
        \begin{center}
          \includegraphics[clip,width=0.99\hsize]{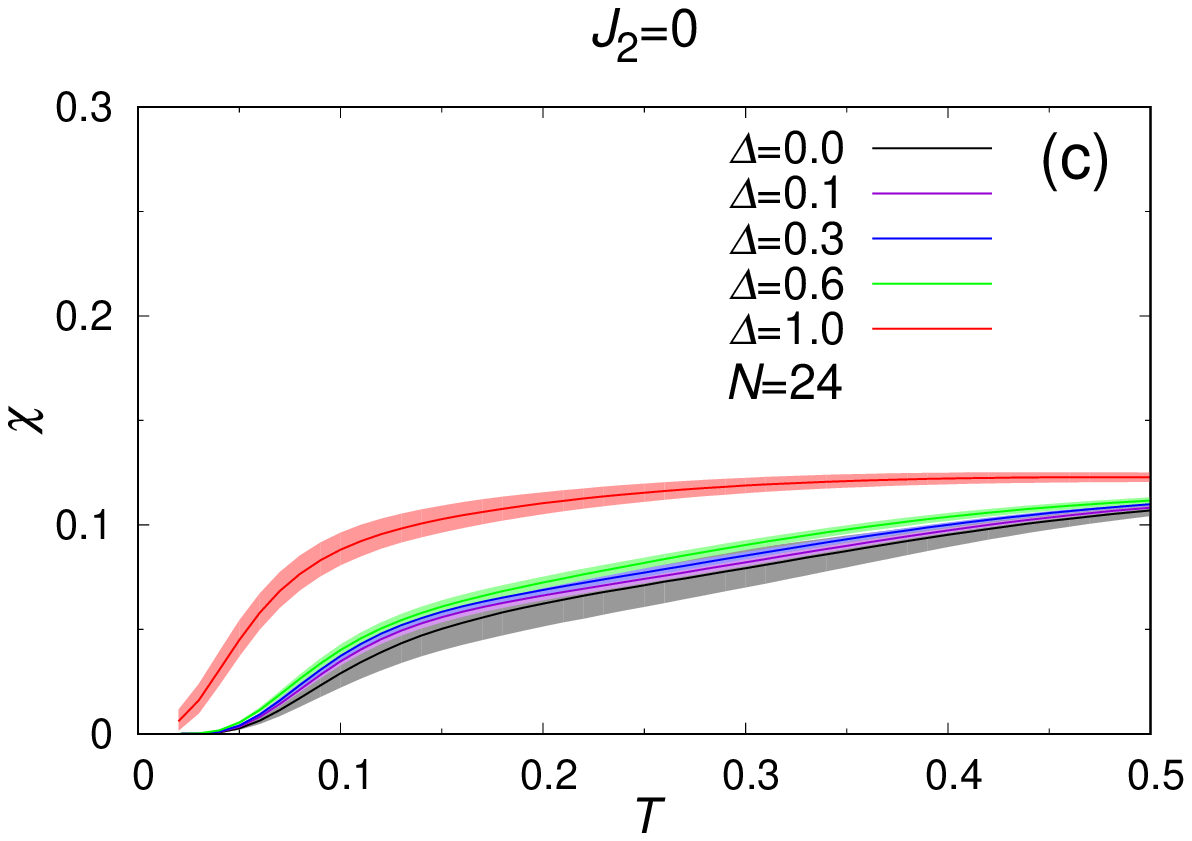}
        \end{center}
      \end{minipage}
      \begin{minipage}{0.5\hsize}
        \begin{center}
          \includegraphics[clip,width=0.99\hsize]{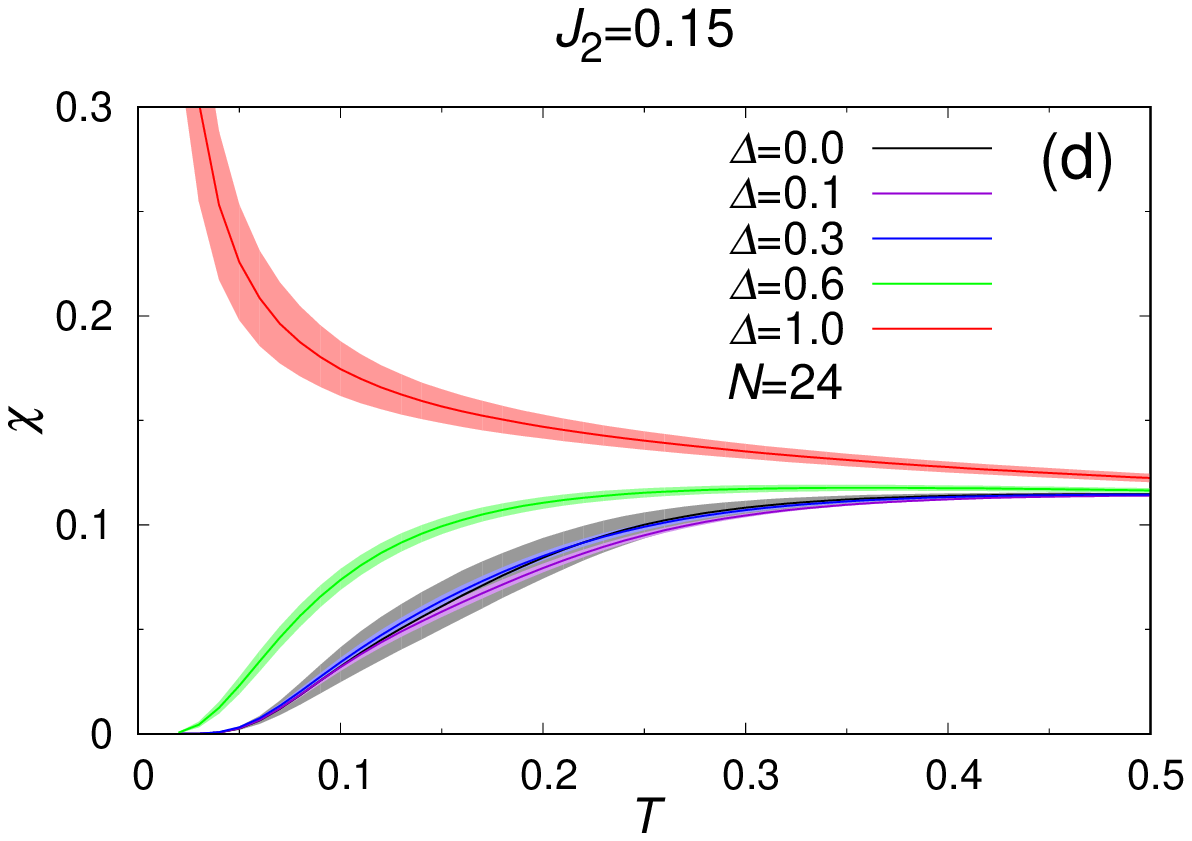}
        \end{center}
      \end{minipage}
    \end{tabular}
    \caption{[Upper row] Temperature dependence of the specific heat per spin $C$ for various values of $\Delta$, at (a) $J_2=0$ and (b) $J_2=0.15$. [Lower row] Temperature dependence of the uniform susceptibility per spin $\chi$ for various values of $\Delta$ at (c) $J_2=0$ and (d) $J_2=0.15$.}
    \label{fig:del--finiT1}
  \end{center}
\end{figure}

 In Fig. \ref{fig:del--finiT1}, we show the temperature dependence of the specific heat per spin for (a) $J_2=0$ and (b) $J_2=0.15$, and of the susceptibility per spin for (c) $J_2=0$ and (d) 0.15.

 As can be seen from Fig. \ref{fig:del--finiT1}(b), in the region of the random-singlet state, {\it e.g.\/}, at $J_2=0.15$ and $\Delta=1$, the computed low-temperature specific heat exhibits a $T$-linear behavior as generically expected for the random-singlet state,\cite{Watanabe,Kawamura} while it exhibits a stronger curvature in the AF state. Likewise, as can be seen from Fig. \ref{fig:del--finiT1}(d), the susceptibility tends to exhibit a gapless behavior with a Curie tail in the region of the random-singlet state, just as expected for that state.\cite{Watanabe,Kawamura} Hence, our finite-temperature data also supports the identification that the ground state at $J_2=0.15$ and $\Delta=1$ is the random-singlet state.

\subsection{Region $0.25\leq J_2\leq 0.5$}
\begin{figure}
  \begin{center}
    \begin{tabular}{c}
      \begin{minipage}{0.5\hsize}
        \begin{center}
          \includegraphics[clip,width=0.99\hsize]{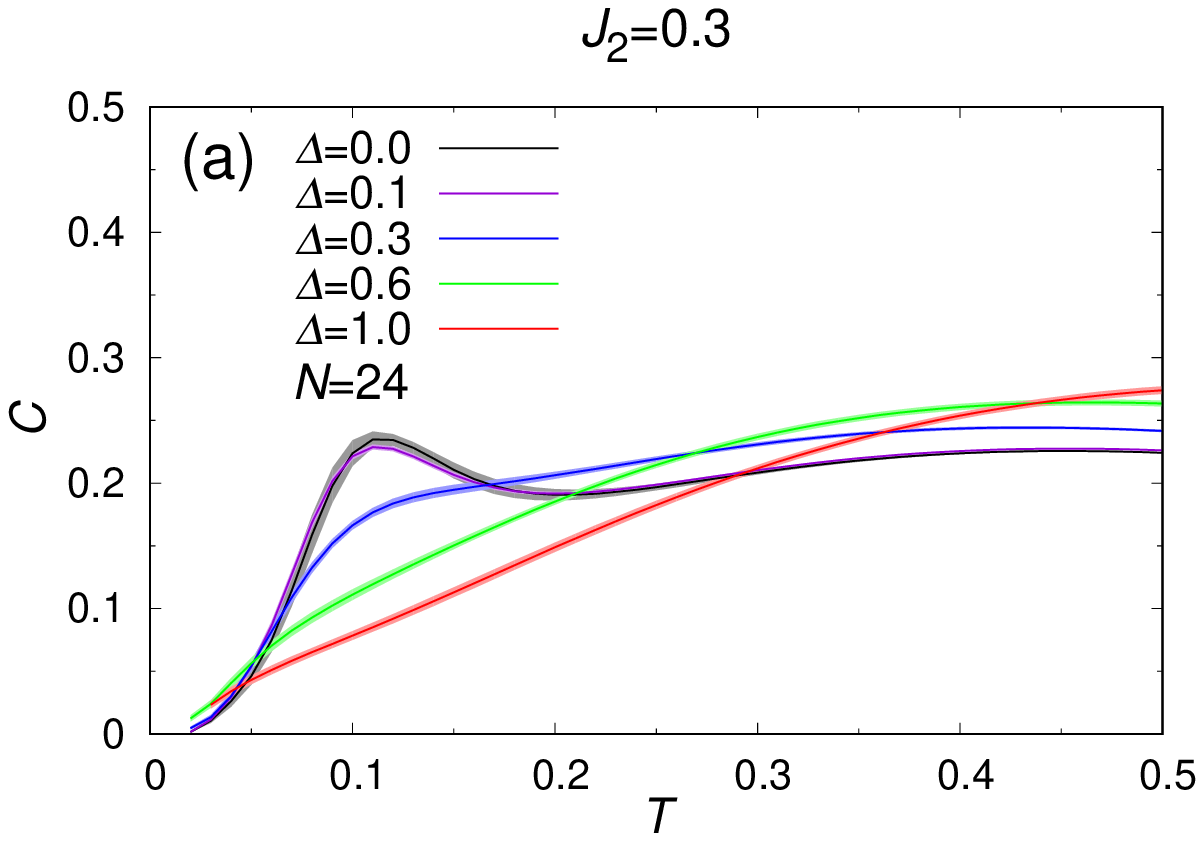}
        \end{center}
      \end{minipage}
      \begin{minipage}{0.5\hsize}
        \begin{center}
          \includegraphics[clip,width=0.99\hsize]{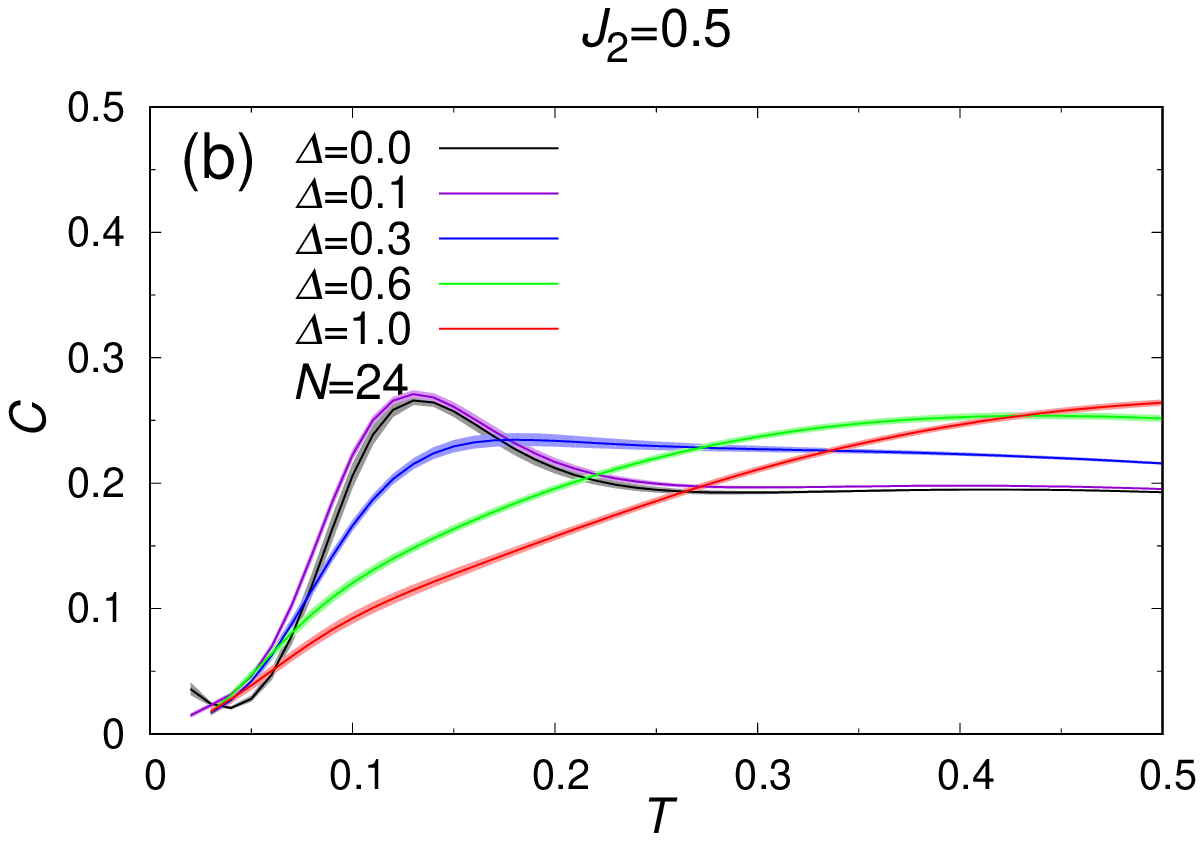}
          \end{center}
      \end{minipage} \\
      \begin{minipage}{0.5\hsize}
        \begin{center}
          \includegraphics[clip,width=0.99\hsize]{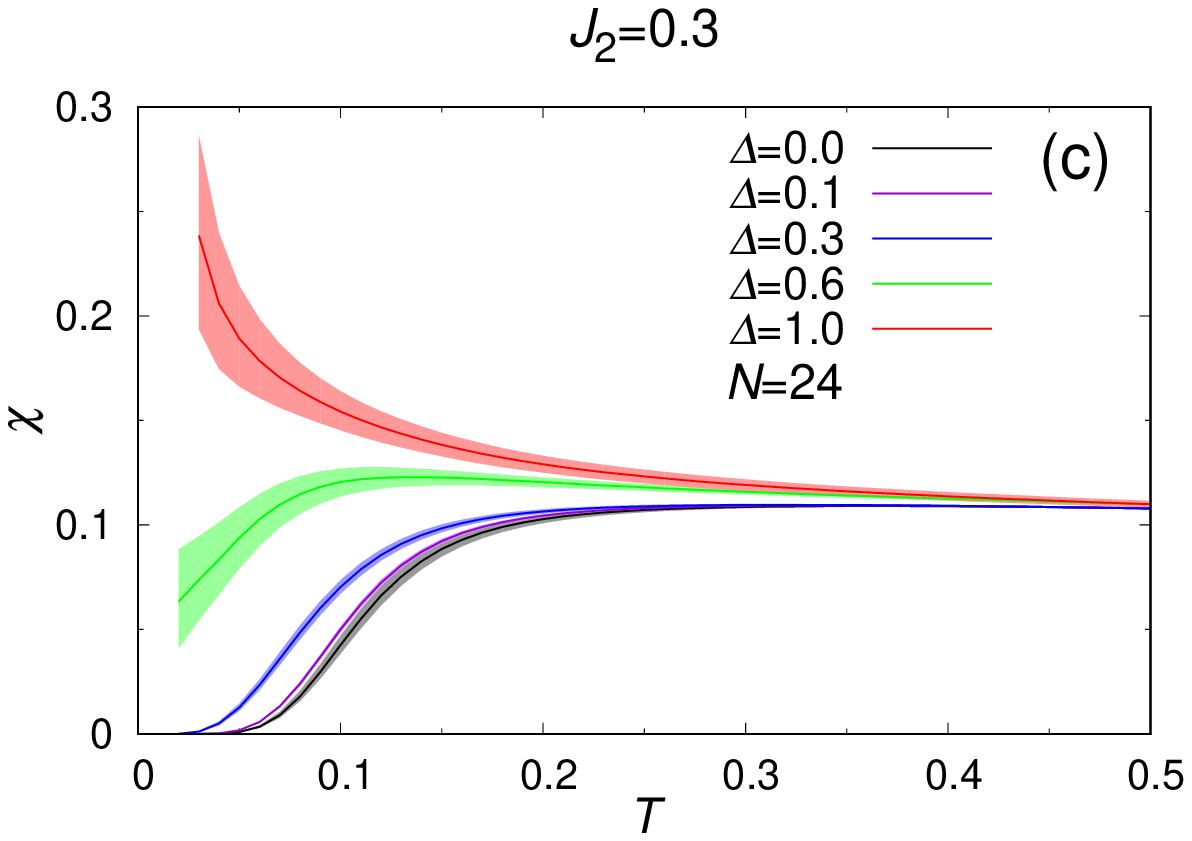}
        \end{center}
      \end{minipage}
      \begin{minipage}{0.5\hsize}
        \begin{center}
          \includegraphics[clip,width=0.99\hsize]{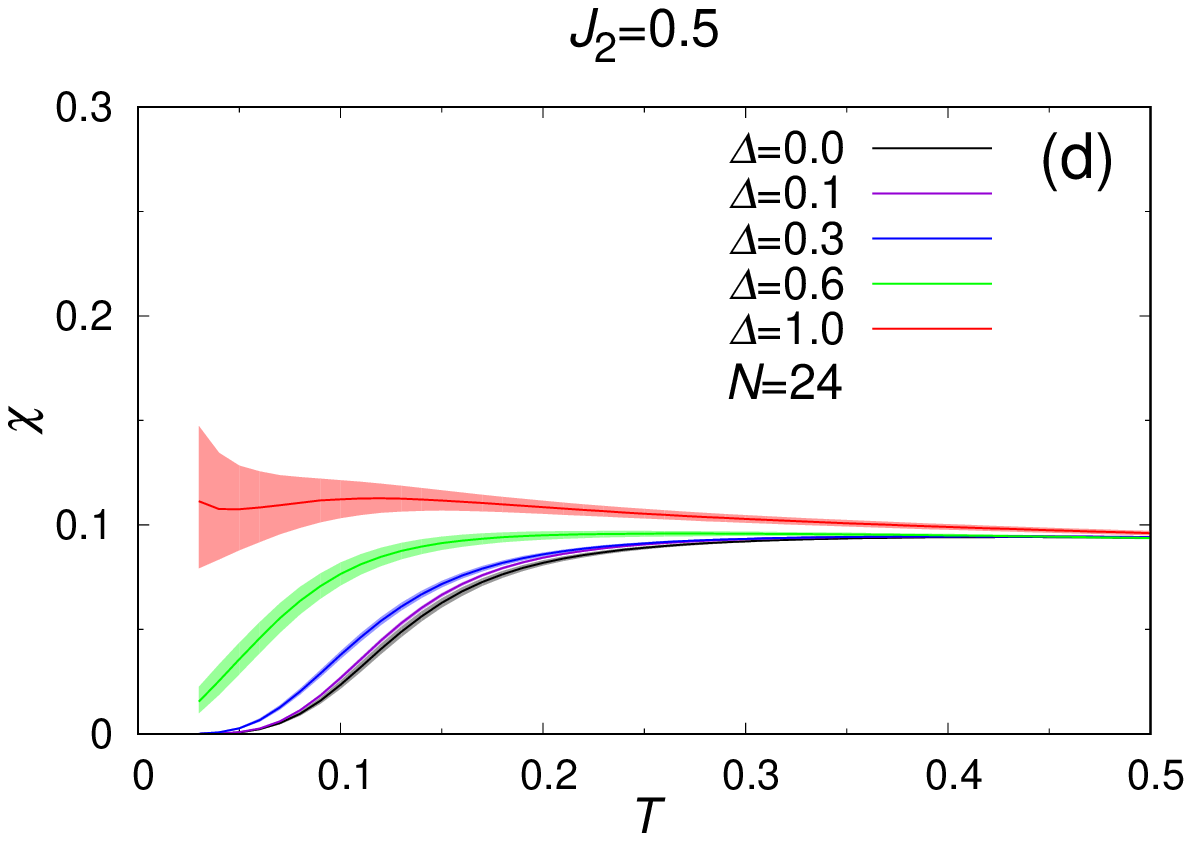}
        \end{center}
      \end{minipage}
    \end{tabular}
    \caption{[Upper row] Temperature dependence of the specific heat per spin $C$ for various values of $\Delta$, at (a) $J_2=0.3$ and (b) $J_2=0.5$.
      [Lower row] Temperature dependence of the uniform susceptibility per spin $\chi$ for various values of $\Delta$ at (c) $J_2=0.3$ and (d) $J_2=0.5$.}
    \label{fig:del--finiT2}
  \end{center}
\end{figure}

Next, we turn to the region $0.25\leq J_2\leq 0.5$. In Fig. \ref{fig:del--finiT2}, we show the temperature dependence of the specific heat for (a) $J_2=0.3$ and (b) $J_2=0.5$, and of the susceptibility for (c) $J_2=0.3$ and (d) $J_2=0.5$.

 As can be seen from Fig. \ref{fig:del--finiT2}(a) and (b), in the region of the random-singlet state, {\it e.g.\/}, at $\Delta=0.6$ and 1 for both $J_2=0.3$ and 0.5, the computed low-$T$ specific heat tends to exhibit a $T$-linear behavior as expected for the random-singlet state.\cite{Watanabe,Kawamura}

 As mentioned in sect. \ref{sec:del00}, the gapped I and II phases stabilized in this regime for weaker randomness are likely to be the plaquette and the staggered-dimer states, each of which spontaneously breaks the $Z_3$ symmetry, the lattice translation in the former and the lattice rotational in the latter. In 2D, a spontaneously breaking $Z_3$ symmetry might accompany a finite-$T$ transition with a divergent specific heat, possibly lying in the universality class of the three-state Potts model, as was pointed out in Ref. \cite{Mulder} for the latter case. As can be seen from Figs. \ref{fig:del--finiT2}(a) and \ref{fig:del--finiT2}(b), the computed specific heat exhibits a double-peak structure in this region, where the lower-temperature peak might correspond to such a finite-temperature $Z_3$-symmetry-breaking transition. In that case, the low-temperature peak would diverge in the thermodynamic limit. Unfortunately, our lattice size $N=24$ presently available is too small to prove or disprove such a theoretical expectation.

 As can be seen from Figs. \ref{fig:del--finiT2}(c) and \ref{fig:del--finiT2}(d), the susceptibility in the region of the random-singlet state tends to exhibit a gapless behavior with a Curie tail. \cite{Watanabe,Kawamura} This also lends support to our identification of the random-singlet state in the phase diagram. In Fig. \ref{fig:del--finiT2}(d), the Curie tail for $\Delta=1$ is very weak and is hardly visible in the temperature range studied. However, a finite fraction of samples turns out to have triplet ground states, suggesting that a weak Curie tail would eventually show up at sufficiently low temperatures even in this case. 


\section{Summary and discussion}
\label{sec:summary}

 Both the ground state and the finite-temperature properties of the random-bond $s=1/2$ $J_1-J_2$ Heisenberg model on the honeycomb lattice are investigated by the ED and the Hams--de Raedt methods. The ground-state phase diagram is constructed in the randomness ($\Delta$) versus the frustration ($J_1/J_2$) plane in order to obtain insight into the role of randomness and frustration in stabilizing various phases. Without frustration, {\it i.e.\/}, for $J_2=0$, the AF LRO is kept stable up to the maximal randomness of $\Delta=1$. Hence, frustration plays a role in destabilizing the AF order. In other words, just the randomness is insufficient to induce the random-singlet state. In the phase diagram, we found three types of nonmagnetic states stabilized. For the regular and the weakly random cases, we found, on increasing $J_2$, the gapped I and gapped II phases, each likely to be the plaquette and the staggered-dimer phases, respectively. For the case of stronger randomness, we generically found a randomness-induced gapless QSL-like state, a random-singlet state, essentially of the same type as previously identified for the random triangular and kagome models. The observed robustness of the random-singlet state suggests that the gapless QSL-like behaviors might be realized in a wide class of frustrated quantum magnets possessing a certain amount of randomness or inhomogeneity, without fine-tuning the interaction parameters.

 Note that we discussed the random-singlet state, of the type where the high symmetry of the underlying lattice is observed at the {\it macroscopic\/} level, even though each $J_{ij}$ realization {\it microscopically\/} breaks the high symmetry of the lattice. In some other situations, the lattice symmetry might be lowered even at the macroscopic level via, {\it e.g.\/}, the possible uniform JT distortion as in the case of orthorhombic samples of Ba$_3$CuSb$_2$O$_9$. Such a symmetry lowering would often enhance the spatially aligned singlet formation such as the VBC, and tends to induce a finite spin gap, {\it i.e.\/}, it tends to induce the gapped nonmagnetic state rather than the gapless nonmagnetic state. 

 On the basis of these findings, we now wish to discuss the possible experimental implications of our present results, particularly regarding the honeycomb-lattice-based magnets 6HB-Ba$_3$NiSb$_2$O$_9$ and Ba$_3$CuSb$_2$O$_9$. First, we wish to discuss the QSL-like behavior recently observed in the $s=1$ honeycomb AF 6HB-Ba$_3$NiSb$_2$O$_9$. \cite{Cheng,Mendels-Ni,Darie} This compound is found to exhibit gapless QSL-like behaviors, accompanied by the $T$-linear low-$T$ specific heat and the gapless susceptibility with a Curie tail. As mentioned in Ref. \cite{Mendels-Ni}, the exchange interaction in this material might be of the $J_1$-$J_2$ honeycomb-type, where a considerable amount of structural disorder exists. The upper limit of the exchange disorder was estimated to be $\Delta\lesssim 0.25$. \cite{Mendels-Ni} Although the ratio $J_2/J_1$ is not known precisely, the structure of this compound suggests a moderately large $J_2$. Of course, the $s=1$ nature of this compound might somewhat modify the phase diagram of the $s=1/2$ model obtained here. For example, in the case of the regular model, the $J_2$-value at the AF-plaquette phase boundary was estimated to be $J_{2c}\simeq0.3$, \cite{Gong-s=1, Li-s=1} which should be compared with the corresponding estimate $J_{2c}\simeq 0.25$ for the $s=1/2$ model. Even with such an uncertainty, the existence of a significant amount of exchange randomness of $\Delta \simeq 0.25$ is likely to locate 6HB-Ba$_3$NiSb$_2$O$_9$ lying in the random-singlet state or close to its phase boundary: see Fig. \ref{fig:phasediagram-random}. Hence, a good possibility seems to exist that the experimentally observed gapless QSL-like behavior of 6HB-Ba$_3$NiSb$_2$O$_9$ might indeed be that of the random-singlet state. In fact, in Ref. \cite{Mendels-Ni}, within the modeling of 6HB-Ba$_3$NiSb$_2$O$_9$ by the random {\it triangular\/} model, the possibility of the random-singlet state as proposed in Ref. \cite{Watanabe} for the random triangular model was examined, but led to a negative result, arguing that the extent of the randomness deduced for 6HB-Ba$_3$NiSb$_2$O$_9$ is smaller than the critical randomness of the random triangular model, $\Delta_c\simeq 0.5$. Yet, as mentioned above, this compound is likely to be better modeled as the random $J_1-J_2$ {\it honeycomb\/} model with moderately large $J_2$, to which our present study suggests $\Delta_c\simeq 0.25$ much smaller than the corresponding value of the random triangular model. Hence, the modeling of 6HB-Ba$_3$NiSb$_2$O$_9$ as the $J_1$-$J_2$ honeycomb model leaves a good possibility of the random-singlet state. Of course, care has to be taken in a truly quantitative comparison, because the type of randomness assumed here is simplified in that it obeys a simple uniform distribution and is taken to be completely bond-independent.

For the hexagonal sample of Ba$_3$CuSb$_2$O$_9$, in Ref. \cite{Nakatsuji}, two scenarios were proposed,  {\it i.e.\/}, a {\it random static\/} JT-distortion-driven `random-singlet state' versus a {\it dynamical\/} JT-distortion-driven `spin-orbital liquid state'. The former would essentially be of the same character as the random-singlet state discussed in the present paper. Meanwhile, many recent experimental studies point to the second scenario, although the NMR study \cite{Mendels-Cu} and the density-functional calculation \cite{Shanavas} suggested the first scenario. If the second scenario applies, to properly understand the QSL-like behaviors observed in the hexagonal sample of Ba$_3$CuSb$_2$O$_9$, one would need to consider the effect of fluctuating orbital degrees of freedom. \cite{Ishiguro,Hagiwara,Katayama,Nasu,Mila,Nasu2}

 For the orthorhombic sample, on the other hand, a {\it symmetry-lowering static\/} JT distortion occurring at a higher temperature might somewhat modify the nature of the low-temperature spin state (also called `random-singlet state' in Ref. \cite{Do}) from the one discussed in the present paper, with an enhanced character of the VBC and a finite spin gap. Quantitative details of the gapless/gapful issue would depend on the competition between the extent of the randomness and that of the uniform distortion in $J_{ij}$, and needs further clarification. 

 In any case, the randomness-induced gapless QSL-like state, the random-singlet state, prevails in quantum magnets on a variety of frustrated lattices, including not only the $J_1-J_2$ honeycomb lattice as studied here, but also the triangular and the kagome lattices, if a certain amount of randomness or inhomogeneity is introduced in some way or another. This would further extend our concept of the QSL state as a novel state of matter. 

 \section*{Acknowledgements}
 The authors wish to thank T. Shimokawa and T. Okubo for valuable discussion. This study was supported by JSPS KAKENHI Grant Number JP25247064. Our code was based on TITPACK Ver.2 coded by H. Nishimori. We are thankful to ISSP, the University of Tokyo, and to YITP, Kyoto University, for providing us with CPU time.

\end{document}